\documentclass[twocolumn]{autart}
\usepackage{graphicx}
\usepackage{color}
\usepackage{amsmath, amssymb}
\usepackage{cite}
\usepackage{algorithm}
\usepackage{algpseudocode}
\usepackage{latexsym}
\usepackage{hyperref}
\usepackage[usenames,dvipsnames,svgnames,table]{xcolor}
\definecolor{mypink2}{RGB}{219, 48, 122}
\usepackage{wrapfig}
\usepackage{wasysym}
\usepackage{soul}
 
\newtheorem{lemma}{Lemma}
\newtheorem{remark}{Remark}
\newtheorem{assumption}{Assumption}

\newcommand{\Om}{(\omega)}
\newcommand{\Se}{\hat{S}}
\newcommand{\V}{\mathbb{V}}
\newcommand{\E}{\mathbb{E}}

\begin{document}
\begin{frontmatter}

\title{MIMO ILC for Precision SEA Robots using Input-Weighted Complex-Kernel Regression\thanksref{footnoteinfo}} 

\thanks[footnoteinfo]{This research was conducted in the Mechanical Engineering Department at the University of Washington. 
}

\author{Leon Yan}\ead{liangy00@uw.edu},  \author[]{Nathan Banka}\ead{nathan.banka@gmail.com},  
\author{Parker Owan}\ead{mrpowan10@gmail.com}, \author[]{Walter Tony Piaskowy}\ead{apiasko2@gmail.com},
\author{Joseph Garbini}\ead{garbini@uw.edu},
\author{Santosh Devasia}\ead{devasia@uw.edu}


\begin{abstract}                          
This work improves the positioning precision of lightweight robots with series elastic actuators (SEAs). Lightweight SEA robots, along with low-impedance control, can maneuver without causing damage in uncertain, confined spaces such as inside an aircraft wing during aircraft assembly. Nevertheless, substantial modeling uncertainties in SEA robots reduce the precision achieved by model-based  approaches such as inversion-based feedforward. Therefore, this article improves the precision of SEA robots around specified operating points, through a multi-input multi-output (MIMO), iterative learning control (ILC) approach. The main contributions of this article are to (i)~introduce an input-weighted complex kernel to estimate local MIMO models using complex Gaussian process regression (c-GPR)  (ii)~develop
Geršgorin-theorem-based conditions on the iteration gains for ensuring ILC convergence to precision within noise-related limits, even with errors in the estimated model; and (iii)~demonstrate precision positioning with an experimental SEA robot. Comparative experimental results, with and without ILC, show around $90$\% improvement in the positioning precision (close to the repeatability limit of the robot) and a $10$-times increase in the SEA robot's operating speed with the use of the MIMO ILC.
\end{abstract}

\end{frontmatter}

\section{Introduction}
Lightweight, relatively-small, SEA robots  are well suited for automating manufacturing tasks in confined spaces such as an aircraft wing or tail, e.g., for cleaning pilot holes (used for fixtures) during aircraft assembly. Often these are hard to reach spaces, and workers need to  crawl through tight spaces and manual operations  can be ergonomically challenging. The  robot needs to be lightweight 
(around 20 pounds or less) for easy placement, often by hand. Moreover, direct estimates of the joint forces found by measuring the deformation of the elastic element in SEAs~\cite{Pratt1995}, 
along with low-impedance control can limit the forces applied by the robot when maneuvering under uncertainty in  confined spaces, which 
can help avoid potential damage to the workpiece and costly repairs. 
Nevertheless,  this increased control over forces during maneuvering  in uncertain environments comes at the 
cost of lower  positioning precision of lightweight robots~\cite{Paine2014}, even at the final operating points where the environmental uncertainty might be lower.  However, precision positioning, around an operating point is desirable since it can enable the use of  lightweight SEA robots in manufacturing tasks  in hard-to-reach, confined spaces. Such precision in conjunction with  compliance (which avoids damage)  can be especially beneficial for contact-type applications such as cleaning of holes.  
The problem in achieving precision is that the flexural systems in lightweight SEA robots result in   non-minimum phase dynamics, 
and high gains (for improved performance) can lead to instability.  Therefore, there are limits to the achievable precision with feedback~\cite{davison93}. 
In general,  model-based methods, such as inversion to find feedforward inputs to augment the 
feedback, can be used to improve  performance of  robots, 
e.g.,~\cite{Spong_87}.
A challenge to accurate modeling is the difficulty in capturing nonlinearities and contact-related effects in SEA robots, e.g.,~\cite{Eppinger1988}.
Therefore, tracking error with such model-inversion-based approaches can be large if there are significant model estimation errors.
This motivates the current work  to improve precision-positioning  of SEA robots at different operating points (around which the dynamics can be linearzied), even in the presence of  modeling uncertainties. 
While the current approach is shown to be well suited to improve precision around an operating point, it is not suited for applications where large robot motions are required. In such cases, the nonlinearity can be significant and alternative approaches such as  Lyapunov-based iterative methods~\cite{altin2017exponential} or iterative methods for model-based impedance control~\cite{li2018iterative} could be considered. Moreover, when the signals are periodic and band-limited, the frequency-domain regression method developed in the current article  could be extended to the nonlinear case using generalized frequency-response functions as in~\cite{stoddard2019gaussian}.

Iterative learning methods can enable precision control even in the presence of 
modeling uncertainties, especially when the task can be learned ahead of time or can be 
repeated~\cite{arimoto_bettering_84}.
Moreover, previously learned precision sub-tasks could be recombined to form new tasks~\cite{Sandipan_tomi}.
When the model uncertainty is low,  model inversion can be used to improve the performance of robots with elastic joints, e.g.,~\cite{Lucibella_98} and can be applied to correct for nonminimum-phase flexural 
dynamics of lightweight robots~\cite{PaChLeBa}. 
Therefore, in this article, a model-inversion-based ILC~\cite{jayati00,ye2005clean,tien2005iterative} is used to account for modeling uncertainties with the lightweight SEA robot. 
However,  the ILC iterations could potentially diverge if the modeling uncertainty is large, e.g.,~\cite{Moore07_review}. This motivates approaches developed to reduce the model uncertainty to 
improve convergence~\cite{kim2012modeling}, by using observed input-output data ($I_O\Om, O_O\Om$ at frequency $\omega$) of the form ${I_O\Om}/{O_O\Om}$ for single-input single-output (SISO) systems, rather than using the inverse of a known model. 
Even with the direct use of input-output data, the effective model uncertainty $\Delta\Om$ can 
be large still if the  signal to noise ratio 
is small, e.g., when the desired output $O_O\Om$ at some frequency $\omega$ is small~\cite{deutschmann2018modeling}. An approach is to not iterate at those specific
frequencies where the desired output is small, e.g.,~\cite{kim2012modeling,de2019data}.
Another approach is to inject additional input at such frequencies (where the desired output is small) to ensure 
persistence of excitation when estimating the model from data, which enables the learned model to be portable and applicable to track new trajectories, e.g.,~\cite{devasia2017iterative}.
In either case, provided the uncertainty is sufficiently small,  the ILC converges to the input needed for exact output tracking  for SISO systems,
even in the presence of modeling uncertainties. 
Therefore, it is expected that  the model-inversion-based 
ILC approach 
could improve the precision  for the MIMO case, in the presence of small uncertainties, e.g., with the use of a linearized model of the SEA robot obtained at an operating point. 

While convergence conditions have been well established for the SISO inversion-based ILC~\cite{tien2005iterative}, extension to the MIMO inversion-based ILC remains challenging. For sufficiently-diagonally-dominant systems, convergence can be established  with the use of a diagonalized inverse model in the ILC as shown in ~\cite{yan2012decoupled}. Recent efforts~\cite{banka2018iterative,de2019multivariable} have considered ILC with the full MIMO inverse~\cite{ye2005clean}. 
For example, the  ILC gain $\rho\Om$ could be optimized to minimize the tracking error  with the constraint that the iterations converge, e.g., by using a Q-filter when computing the ILC input~\cite{de2019multivariable}. The Q-filter approach  trades off between robustness and performance; if the anticipated uncertainty is small, then the tracking error is small, but not zero, when the Q-filter is nontrivial, i.e., $Q\Om \ne 1$. In contrast, this article develops  conditions on the ILC gain,  using the Geršgorin theorem as in~\cite{banka2018iterative}, for ensuring convergence to the desired output, even under  modeling uncertainty.

A challenge with inversion-based ILC is that the actual system dynamics $S\Om$ and the uncertainty $\Delta\Om$ are unknown. Therefore, it is difficult to ensure that the convergence condition on the model uncertainty is met. One approach is to use  a relatively large number of repetitive experiments for identifying both the model and the uncertainty prior to applying the ILC~\cite{tien2005iterative,de2019multivariable}. 
An alternative approach is to use 
kernel-based regression approaches to estimate the model and its uncertainty for the SISO case~\cite{devasia2017iterative} from data obtained from a relatively-sparse  number (one or two) of the ILC iterations. 
This use of ILC data allows the data-based model to capture the local, linear model near the work area that includes potential contact-dependent effects,
which can be challenging to model from first principles. 
Note that kernel-based Gaussian process regression (GPR, popularly also referred to as machine learning) is well suited to estimation of the general functions (such as the  frequency response function) as well as their uncertainty from  noisy experimental  data~\cite{Rasmussen_06}.
Though complex kernels have been used in the past for GPR~\cite{boloix2018complex}, they  cannot be used to estimate models for multi-input systems.
This motivates the development, in the current article, of an input-weighted complex kernel to estimate the complex-valued system MIMO model $\hat{S}\Om$ as well as its uncertainty $\Delta\Om$. 
Since it is designed in the Fourier domain, it captures the noncausality needed for many system inversion approaches, e.g., as investigated in~\cite{blanken2020kernel},  and can be used with SISO frequency-domain kernels which ensure BIBO stability of the resulting models, e.g., as in~\cite{lataire2016transfer}. 
Such kernel-based methods tend to be robust than typical  model-identification methods in infinite-dimensional 
spaces, e.g., as discussed in~\cite{pillonetto2014kernel}. 

\vspace{-0.1in}
The current work extends preliminary results in \cite{banka2018iterative} by  (i)~proving the conditions for convergence, 
(ii)~using augmented inputs in the ILC to ensure sufficiently large signal-to-noise ratio in the frequency range of interest, and
(iii)~applying and evaluating the approach for precision control of a lightweight SEA robot.
Comparative experimental results, with and without ILC, are presented that show  around $90$\% improvement in the positioning precision 
with the use of the MIMO ILC 
(close to the repeatability limit of the robot)
and a $10$-times increase in the SEA robot's operating speed. 

\vspace{-0.1in}

\section{Problem Formulation}
\vspace{-0.1in}
\subsection{Inversion-based iterative-learning control (ILC)}
Given a linear time-invariant (LTI) system with transfer function (or frequency response function) $S\Om$ 
\begin{equation}
	O\Om = S\Om I\Om,
    \label{eq_model_freq_response_abstract}
\end{equation}
with output $O\Om$ and input $I(\omega)$ represented in the Fourier domain at frequency $\omega$, 
the goal  is to find an input 
$I_d\Om$ that yields exact tracking of  the desired output $O_d\Om$, i.e., 
\begin{equation}
	O_d\Om = S\Om I_d\Om,
    \label{eq_desired_response}
\end{equation}
where the system $S\Om$ is stable and the number of outputs $m$ is not more than  the number of inputs $n$, i.e., $m \le n$.

\begin{assumption}
\label{exact_tracking_existence}
The LTI system $S\Om \in \mathbb{C}^{m\times n}$  has full row rank on the imaginary axis. 
\end{assumption}

\begin{remark}
The rank condition implies that there are no transmission zeros on the imaginary axis, which 
guarantees the existence of a finite input $I_d\Om$ that achieves exact tracking of the desired output $O_d\Om$
and 
ensures  robustness of the inverse~\cite{devasia_02_tac}. However, conditions have been studied recently on when such zeros are acceptable for inversion~\cite{naderi2018inversion}.
\end{remark}

The iterative approach generates a new input $I_{k+1}$ for use at iteration step $k+1$ based on the tracking error 
\begin{equation}
E_{k}\Om = O_d\Om -O_{k}\Om
    \label{eq_tracking_error}
\end{equation}
from the prior iteration step $k$, which is similar to prior use for the square system case, e.g., in~\cite{tien2005iterative,ye2005clean,banka2018iterative}  %
\begin{equation}
I_{k+1}\Om  =  I_{k}\Om + \Se^{\dagger}\Om\rho\Om(O_d\Om - O_k\Om) \label{eq_update_IIC}
\end{equation}
where
$ \Se^{\dagger}\Om = \textbf{W}^{-1}\Om\Se^*\Om (\Se\Om\textbf{W}^{-1}\Om\Se^*\Om)^{-1}   $
is the pseudo-inverse of the 
estimated model $\Se\Om$ of the unknown system $S\Om$ 
with $\Se\Om \Se^{\dagger}\Om  =  \textbf{1} $ 
 and  the identity matrix is denoted by $\mathbf{1}$. 
The 
superscript $*$ denotes the complex conjugate, $\textbf{W}\Om\in \mathbb{R}^{n\times n}$ is
an invertible diagonal matrix, with $\textbf{W}_{i,i}\Om=~w_i\Om >0$, and  $\rho\Om \in \mathbb{R}^{m\times m}$ is  a diagonal matrix with nonnegative elements representing the iteration gain.

\begin{remark}
For  square systems with the same number of inputs $n$ and outputs $m$, i.e.,  $n=m$, the pseudo-inverse $\Se^{\dagger}\Om$ becomes the exact inverse $\Se^{-1}\Om$.
For the actuator redundant case, i.e., $n >m$, 
$ I\Om = \Se^{\dagger}\Om O\Om $ solves the input-output equation 
$  O\Om  = \Se\Om  I\Om$  while minimizing the frequency-weighted input cost 
$\|I\Om \|_{\textbf{W}\Om}^2 = 
I^*\Om \textbf{W}\Om I\Om
$.
\end{remark}

\noindent 
Multiplying $S\Om$ on both sides of Eq~\eqref{eq_update_IIC} results in
\begin{equation}
O_{k+1}\Om  =  O_{k}\Om + S\Om\Se^{\dagger}\Om\rho\Om(O_d\Om - O_k\Om).
\nonumber
\end{equation}
Subtracting the desired output $O_d\Om$ from both sides yields a relation between error defined in Eq.~\eqref{eq_tracking_error} at consecutive iteration steps, as 

\vspace{-0.3in}
{\small{
\begin{equation}
E_{k+1}\Om  
=  [\textbf{1}-\rho\Om-\Delta\Om\rho\Om] E_{k}\Om= G\Om E_k\Om,
\label{eq_err_contraction}
\end{equation}
}}

\vspace{-0.3in}
\noindent 
where the unknown model uncertainty $\Delta\Om \in \mathbb{C}^{m\times m}$
\begin{equation}
\Delta\Om  = S\Om \Se^{\dagger}\Om -\mathbf{1}.
\label{eq_uncertainty_def}
\end{equation}
In the following, the initial iteration input $I_0$ is selected as the desired output, $I_0\Om = O_d\Om$, which is assumed to be sufficiently smooth and bounded over the frequency range.

\begin{remark}
The model uncertainty $\Delta\Om$ in Eq.~\eqref{eq_uncertainty_def} 
quantifies the error in the estimated model $\Se$, or equivalently the error in computing the pseudo-inverse $\Se^{\dagger}$. 
\end{remark}

\begin{remark}
From Assumption~\ref{exact_tracking_existence}, the 
estimated  model $\Se\Om $  has full row rank on the imaginary axis for sufficiently small model
estimation error 
 $\delta\Om = S\Om -\Se\Om $. 
\end{remark}

\subsection{Problem statement}
\label{sec:res_prob}
The iterations result in a contraction if the model uncertainty $\Delta\Om$ is sufficiently small. 
From Eq.~\eqref{eq_err_contraction}, the tracking error 
$E_{k+1}=[G\Om]^{k} E_1\Om$,
and therefore, with increasing iterations, the tracking error $E\Om$ tends to zero in the 2-norm, if and only if
$\lim_{k\rightarrow\infty}
[G\Om]^k = \mathbf{0}$, 
which in turn occurs if and only if
the spectral radius $\sigma_G\Om $  of the contraction gain $G\Om$ is less than one as in Eq.~\eqref{eq_contraction_gain_condition}, e.g.,~see Theorem 5.6.12 in \cite{horn2012matrix}.
Therefore, as shown in the previous works e.g.,~\cite{tien2005iterative,ye2005clean,banka2018iterative}, the MIMO ILC in Eq.~\eqref{eq_update_IIC} converges at frequency $\omega$ for any general output $O_d\Om$, 
i.e., the tracking error tends to zero
\begin{equation}
\lim_{k\rightarrow \infty}
\|E_k\Om \|_2  = 0
\label{eq_meaning_convergence}
\end{equation}
as iterations increase, $k \rightarrow \infty$, if and only if 
the spectral radius $\sigma_G\Om$ (maximum magnitude of the eigenvalues $\lambda_i(G\Om)$ ) of the 
the contraction gain
\begin{align}
G \Om & = [\mathbf{1}-\rho\Om]  -\Delta\Om \rho\Om 
	\label{eq_contraction_gain}
\end{align}
is less than one, i.e., 
\begin{align}
\sigma_G\Om ~ = 
\max_i | \lambda_i(G \Om) |< 1.
	\label{eq_contraction_gain_condition} 
\end{align}

\begin{remark}
\label{equivalance_norm}
Convergence of the tracking error $E_k\Om$ to zero in the two norm also results in convergence in any other norm, since all norms are equivalent in  finite-dimensional linear vector spaces, and for any norm $\| \cdot \|_p$ there exists a constant $\gamma_p$ such that~\cite{horn2012matrix}
$ \|E_k\Om \|_p  \le \gamma_p \| E_k\Om \|_2$.
\end{remark}

The research issue is to develop conditions on the iteration gain $\rho\Om$ to ensure convergence of the ILC under given bounds $\overline{\Delta}\Om$ on the model uncertainty $\Delta\Om$. This can be split into a model and uncertainty estimation problem and the problem of selecting the iteration gain $\rho\Om$  to ensure convergence. 
\begin{enumerate}
    \item Given 
    input-output data ($I_O\Om, O_O\Om$), estimate a model $\hat{S}\Om$ of the system $S\Om$ and  bound $\overline{\Delta}\Om$ on the  model uncertainty $\Delta\Om$ in Eq.~\eqref{eq_uncertainty_def}.
    \item Given a  bound $\overline{\Delta}\Om$ on the model uncertainty $\Delta\Om$, develop  conditions on the iteration gain $\rho\Om$ to  satisfy the ILC convergence condition in Eq.~\eqref{eq_contraction_gain_condition}.
\end{enumerate}

\section{Solution: ILC design for convergence}

\vspace{-0.1in}
\subsection{Model and uncertainty estimation}
The goal is to estimate the model terms $\hat{S}_{j,l}\Om$ from observation $O_{j,O}$ given by 
\begin{equation}
\begin{split}
O_{j,O}(\omega,I_O\Om)&=O_j(\omega,I_O\Om)+\epsilon_j\\
&=\sum_{l=1}^n S^P_{j,l}\Om I_{l,O}\Om+\epsilon_j,
\end{split}
\label{eq_output_MISI_process}
\end{equation}
where $\epsilon_j$ is the measurement noise and 
system components $S_{j,l}\Om$ are replaced by Gaussian Processes $S^P_{j,l}\Om$ 
due to  noise as in previous works, e.g., ~\cite{lataire2016transfer,pillonetto2014kernel}.  

If only one input, say $I_{l}\Om$, is nonzero,  then the system component $S_{j,l}\Om = O_{j}\Om/I_{l}\Om$ is a complex function that can be estimated using Gaussian Process Regression (GPR) 
with previously-developed complex kernels, e.g., the one in~\cite{lataire2016transfer,boloix2018complex,banka2018application}.
However, such an approach requires  $n$ separate experiments with only one nonzero input in each experiment, which might not be feasible in general. Such approaches are not directly applicable to the multi-input case. 

Towards model identification for the multiple-input case, given any set of complex kernels $\hat{k}_l(\cdot,\cdot)$  for $1 \le l \le n $, the following
input-weighted kernel $\hat{k}'(\cdot,\cdot)$ 
is proposed for 
the composite Gaussian process $O_{j}$ in Eq.~\eqref{eq_output_MISI_process}, 
\begin{equation}
\begin{split}
\hat{k}'&([\omega_r,\{I_{l,O}(\omega_r)\}_{l=1}^n] ,[\omega_s,\{I_{l,O}(\omega_s)\}_{l=1}^n])\\ 
&= \sum_{l=1}^nI_{l,O}(\omega_r)\hat{k}_{l}(\omega_r,\omega_s)I^*_{l,O}(\omega_s), 
\end{split}\label{eq_input_weighted_kernel}
\end{equation}
where the subscripts $r,s$ 
refer to different frequencies belonging to the observed frequency set
\begin{equation}
    \Omega=[\omega_1,\dots,\omega_q]\in\mathbb{R}^q
    \label{Omega_Eq}
\end{equation} 
with nonzero accompanying input $\{I_{l,O}(\Omega) \in \mathbb{C}^{q}\}_{l=1}^n$.
A motivation for the proposed kernel is that for the single-input case ($n=1$), using the input-weighted kernel $\hat{k}'$ to
estimate the component $S^P_{j,1}\Om$ 
is equivalent to obtaining an 
estimate with the  unweighted kernel  $\hat{k}_1$ based on observed ratios  $O_{1,O}(\omega)/I_{1,O}(\omega)$, as shown in Lemma~\ref{lem_siso_kernel_equivalence} in the following.
%
Moreover,  $\hat{k}'$
is a valid kernel (Hermetian positive semi-definite) provided the underlying kernels  $\hat{k}_{l}$ associated with each term are valid kernels, as shown in the lemma below under the following assumption.
\begin{assumption}
\label{GPR_assumptions}
All the complex Gaussian Processes $S^P_{j,l}$ are independent, and have zero-mean prior. Moreover they  are assumed to (i)~be proper as in~\cite{schreier2010statistical,boloix2018complex}, resulting in kernel functions $\hat{k}_l$ that have the same variance for the real and imaginary parts at each frequency $\V[\operatorname{Re}(S^P_{j,l}(\omega))]=\V[\operatorname{Im}(S^P_{j,l}(\omega))]=\V_{j,l}\Om$; and (ii)~have independent real and imaginary components resulting in $\V[S^P_{j,l}\Om]=  \V[\operatorname{Re}(S^P_{j,l}(\omega))]
+\V[\operatorname{Im}(S^P_{j,l}(\omega))]=2\V_{j,l}\Om$.
\end{assumption}
\begin{lemma}
\label{rem_mimo_kernel}
With nonzero input $I_O$
at each frequency in the observed frequency set $\Omega$,
the  input-weighted kernel $\hat{k}'$
in Eq.~\eqref{eq_input_weighted_kernel} is 
Hermitian positive semi-definite if the kernels $\hat{k}_l$ $1 \le l \le n$ are Hermitian positive semi-definite. Moreover, the self covariance matrix  $K'$
in Eq.~\eqref{eq_miso_kernel_var}, 
associated with  $\hat{k}'$,
is positive definite if  covariance matrices $K_l(\Omega,\Omega)$ are positive  definite for $1 \le l \le n$.
\end{lemma}

\vspace{-0.2in}
\begin{pf}
The Hermetian property of  the input-weighted kernel $\hat{k}'$ follows from Eq.~\eqref{eq_input_weighted_kernel} since 
\begin{equation*}
\begin{split}
\hat{k}'&([\omega_r,\{I_{l,O}(\omega_r)\}_{l=1}^n] ,[\omega_s,\{I_{l,O}(\omega_s)\}_{l=1}^n])\\ 
&= \sum_{l=1}^nI_{l,O}(\omega_r)\hat{k}_{l}(\omega_r,\omega_s)I^*_{l,O}(\omega_s)\\
&= \sum_{l=1}^n \left\{ I_{l,O}(\omega_s)\hat{k}_{l}(\omega_s,\omega_r)I^*_{l,O}(\omega_r)\right\}^*\\
&=\left\{\hat{k}'([\omega_s,\{I_{l,O}(\omega_s)\}_{l=1}^n],[\omega_r,\{I_{l,O}(\omega_r)\}_{l=1}^n])\right\}^*
\end{split}
\end{equation*}
and positive semi-definiteness follows since each term in the summation in Eq.~\eqref{eq_input_weighted_kernel} is non-negative. 
The covariance $K'$ can be written as, due to independence of terms from Assumption~\ref{GPR_assumptions}, 
\begin{equation}
\begin{split}
&K'([\Omega,\{I_{l,O}(\Omega)\}_{l=1}^n] ,[\Omega,\{I_{l,O}(\Omega)\}_{l=1}^n])\\ 
& \qquad = \sum_{l=1}^n \text{diag}
(I_{l,O}(\Omega))K_l(\Omega,\Omega) \text{diag}(I_{l,O}^*(\Omega)).
\end{split}\label{eq_miso_kernel_var}
\end{equation}
The  $r^{th}$ row and $s^{th}$ column element of 
the covariance matrix $K'\in\mathbb{C}^{q\times q}$ is the evaluation of the input-weighted kernel $\hat{k}'$ on frequency-input pair $[\omega_r,\{I_{l,O}(\omega_r)\}_{l=1}^n]$ and $[\omega_s,\{I_{l,O}(\omega_s)\}_{l=1}^n]$,
which is computed as the input-weighted summation of the  corresponding 
$r^{th}$ row and $s^{th}$ column 
element of each covariance matrix $K_l\in\mathbb{C}^{q\times q}$ computed as  $\hat{k}_l(\omega_r,\omega_s)$. 
For any vector $v \in \mathbb{C}^{q}$,

\begin{equation*}
\begin{split}
&v^* K'([\Omega,I_O(\Omega)], [\Omega,I_O(\Omega)])v\\
& = v^* \sum_{l=1}^n \text{diag}
(I_{l,O}(\Omega))K_l(\Omega,\Omega) \text{diag}(I_{l,O}^*(\Omega))v \\
&=\sum_{l=1}^n  (\text{diag}(I^*_{l,O}(\Omega))v)^*
K_l(\Omega,\Omega) \text{diag}(I_{l,O}^*(\Omega))v\\
&=\sum_{l=1}^n u_l^* K_l(\Omega,\Omega) u_l,
\end{split}
\end{equation*}
where $u_l=\text{diag}(I_{l,O}^*(\Omega))v$ is nonzero if $v$ is nonzero. 
Thus, $K'([\Omega,I_O(\Omega)],[\Omega,I_O(\Omega)])$ is positive (semi-) definite if for all $l$, $K_l(\Omega,\Omega)$ is positive (semi-) definite. 
\qed \end{pf}

\vspace{-0.1in}
\begin{lemma}[Multi-input system identification]
\label{MISO_sys_id_input_weighted_kernel}
The estimate of each term $S^P_{j,l}\Om$ in Eq.~\eqref{eq_output_MISI_process} is given by
\begin{align}
  \Se_{j,l}\Om ~  &= \E[O_j(\omega,e_l)]
  =\E[S^P_{j,l}\Om]
  \nonumber\\
  &=K'_T(K'+ \sigma_{j,\epsilon}^2\textbf{1})^{-1}O_{j,O}([\Omega, \{I_{l,O}(\Omega) \}_{l=1}^n]), 
  \label{E_mimo_cgpr}
\end{align}
with estimated variance $\V[S^P_{j,l}\Om]=\V[O_{j}(\omega,e_l)] $ given by 
\begin{align}
& \V[S^P_{j,l}\Om] =
K'_0 - K'_T(K'+ \sigma_{j,\epsilon}^2\textbf{1})^{-1}(K'_T)^*, 
\label{V_mimo_cgpr}
\end{align}
where $\sigma_{j,\epsilon}$ is the noise variance at $j^{th}$ output,  $e_l$ is an $n$-by-1 vector with one in the $l^{th}$ component and zero elsewhere, and 
\vspace{-0.1in}
\begin{equation}
\begin{array}{rcl}
 K' & = & K'([\Omega,\{I_{l,O}(\Omega)\}_{l=1}^n] ,[\Omega,\{I_{l,O}(\Omega)\}_{l=1}^n]),  \\
 K'_T & = & K'([\omega,e_l],[\Omega,\{I_{l,O}(\Omega)\}_{l=1}^n]),  \\
 K'_0 & = & K'([\omega,e_l],[\omega,e_l]).  
\end{array} \label{eq_cov_matrix_miso_kernel}
\end{equation}
\end{lemma}

\vspace{-0.25in}
\begin{pf}
Since $S^P_{j,l}\Om=O_j(\omega,e_l)$ from Eq.~\eqref{eq_output_MISI_process}, its estimate and variance follows from standard GPR methods, e.g.,~ \cite{boloix2018complex}.
\hfill \qed
\end{pf}

\begin{lemma}
\label{lem_siso_kernel_equivalence}
For SISO case, the estimate of
$S^P_{1}\Om$ and
its variance at any frequency $\omega$ with the original kernel $\hat{k}_1$ based on ratios $O_{1,O}(\Omega,I_{1,O}(\Omega))/I_{1,O}(\Omega)$ are the same as those obtained with input-weighted kernel $\hat{k}'$ based on $O_{1,O}(\Omega,I_{1,O}(\Omega))$, i.e., 
\begin{align}
\hat{S}_{1}\Om = \E[O_1(\omega,1)] & =   
\E[S^P_{1}(\omega)],  \label{Eq_kernel_expectation}
\\
\V[O_1(\omega,1)] &=
\V[S^P_{1}(\omega)] \label{Eq_kernel_var}.
\end{align} 
\end{lemma}
\begin{pf}
When $n=1$, the covariance matrices in Eq.~\eqref{eq_cov_matrix_miso_kernel} with the input-weighted kernel $\hat{k}'$  can be related to the covariance matrices associated with the original kernel $\hat{k}_1$ based on Eq.~\eqref{eq_input_weighted_kernel} and \eqref{eq_miso_kernel_var}, as
\begin{equation}
\begin{array}{rcl}
K' &   =  &  DK_1(\Omega,\Omega)D^*  =  DKD^*, \\
K'_T &  = & D_T K_1(\omega,\Omega)D^* ~ = D_TK_TD^* ~ = K_TD^*, \\
K'_0 &   =&  
D_TK_1(\omega,\omega)D_T^*
~ =  K_1(\omega,\omega) ~ =K_0,
\end{array} 
\label{eq_def_K_Kp} 
\end{equation}
with $D = {diag}(I_{1,O}(\Omega))$ and $D_T =1$. 
Furthermore, the general
error covariance matrix $C_{\epsilon_1} $ for error in Eq.~\eqref{eq_output_MISI_process} is given by 
\begin{align}
C_{\epsilon_1} & =DC_{\epsilon_S}D^*,  
\label{eq_def_C_Cp} 
\end{align}
where $C_{\epsilon_S}$ is the error covariance matrix of the noise in
\begin{equation}
\frac{O_{1,O}(\omega,I_{1,O}\Om)}{I_{1,O}\Om} =S_{1,O}\Om = S_1^P\Om + \epsilon_S\Om,
\label{eq_system_process}
\end{equation}
because by compraing Eq.~\eqref{eq_system_process} and the case of Eq.~\eqref{eq_output_MISI_process} when  $n=1$, which can be written as below,
\begin{equation}
O_{1,O}(\omega,I_{1,O}\Om)=S^P_1\Om I_{1,O} + \epsilon_1,
\end{equation}
the variance relation between the $\epsilon_1$ and $\epsilon_S$ can be obtained as 
\begin{equation}
\begin{split}
\V[\epsilon_1(\Omega)]&=\E[I_{1,O}(\Omega)\epsilon_S(\Omega)\cdot(I_{1,O}(\Omega)\epsilon_S(\Omega))^*]\\&=D\V[\epsilon_S(\Omega)]D^*.
\end{split}
\end{equation}
Then, substituting  Eqs.~\eqref{eq_def_K_Kp} and \eqref{eq_def_C_Cp} into 
equations to generate the estimation of $O_{1}(\omega,1)$
 and using $O_{1,O}(\Omega,I_{1,O}(\Omega))=DS_{1,O}(\Omega)$ from Eq.~\eqref{eq_system_process}  results in
\begin{equation}
\begin{split}
\E[O_{1}(\omega,1)]
&= K'_T(K'+C_{\epsilon_1})^{-1}O_{1,O}(\Omega,I_{1,O}(\Omega))\\
&= K_TD^*(D(K+C_{\epsilon_S})D^*)^{-1}DS_{1,O}(\Omega)\\
&= K_T(K+C_{\epsilon_S})^{-1}S_{1,O}(\Omega)\\ 
&= \E[S^P_1(\omega)],
\end{split}
\nonumber 
\end{equation}
\begin{equation}
\begin{split}
\V[O_1(\omega,1)]
&= K'_0 - K'_T(K'+C_{\epsilon_1})^{-1}(K'_T)^*\\
&= K_0-K_TD^*(D(K+C_{\epsilon_S})D^*)^{-1}DK_T^*\\
&= K_0-K_T(K+C_{\epsilon_S})^{-1}K_T^*\\
&= \mathbb{V}[S^P_1(\omega)],
\end{split} \nonumber 
\end{equation}
which results in the equivalence claim in the lemma.
\qed  \end{pf}

\begin{remark}
\label{rem_uncertain_coeff_bound}
Bounds  $ \overline{\delta}_{j,l}\Om $ on the model estimation error $ \delta_{j,l}\Om = S_{j,l}\Om - \Se_{j,l}\Om$,  between the estimated model $\Se\Om$ and the  unknown system $S\Om$,
can be obtained in terms of the estimated variance 
$\V[S^P_{j,l}\Om] $ in Eq.~\eqref{V_mimo_cgpr}
with 
$\V[S^P_{j,l}\Om] 
= 2\mathbb{V}[Re(S^P_{j,l}\Om)] $ from Assumption~\ref{GPR_assumptions}, 
as
\begin{equation}\label{ineq_bound_mag}
\overline{\delta}_{j,l}\Om 
~=
\gamma_\delta \sqrt{\mathbb{V}[Re(S^P_{j,l}\Om)]}, 
\end{equation}
where the constant $\gamma_\delta$
can be  larger for specifying a bound $\overline{\delta}_{j,l}\Om$  with a higher confidence level.
\end{remark}

\begin{lemma}
\label{Lemma_bounds_uncertainty}
Each component $\Delta_{j,l}\Om$ of the model uncertainty $\Delta\Om$ defined in Eq.~\eqref{eq_uncertainty_def} is bounded  as 
\begin{equation}
\left|\Delta_{j,l}\Om\right|<\sum_{k=1}^n\left| \Se^{\dagger}_{k,l}\Om \right|\overline{\delta}_{j,k} \Om.
\label{eq_bound_uncertainty_computation}
\end{equation}
\end{lemma}

\vspace{-0.25in}
\begin{pf}
This follows by replacing $ S\Om $ in Eq.~\eqref{eq_uncertainty_def} by 
$ S\Om - \Se_{j,l}\Om + \Se_{j,l}\Om$. 
\hfill \qed
\end{pf}

\subsection{Convergence conditions for MIMO ILC}
\label{sec:bSec}

\begin{lemma}[MIMO ILC convergence conditions]
\label{Lemma_convergence_conditions}
Let each component $\Delta_{j,l}\Om$ 
of the  uncertainty  $\Delta\Om$ in Eq.~\eqref{eq_uncertainty_def} have the form 
\begin{equation}
    \Delta_{j,l}\Om = M_{j,l}\Om  e^{\mathbf{i}\Phi_{j,l}\Om} 
    = A_{j,l}\Om + \mathbf{i}B_{j,l}\Om, 
    \label{eq_unknownModelUncertainty_def}
\end{equation}
with  
magnitude $M_{j,l}\Om$ and  phase  $\Phi_{j,l}\Om$ where $\mathbf{i} = \sqrt{-1}$. Also let the iteration gain $\rho\Om$ in Eq.~\eqref{eq_update_IIC} be diagonal, i.e., $\rho\Om  =\text{diag}(\rho_1\Om,\dots,\rho_m\Om)$. Then, the MIMO ILC in Eq.~\eqref{eq_update_IIC} converges at frequency $\omega$ if, for all $1\le i \le m$, 

\vspace{-0.2in}
\begin{align}
R_i\Om   & <  1+A_{i,i}\Om ,
    \label{eq_banka2} 
\\
0 < \rho_i\Om & < 2 \frac{1+A_{i,i}\Om-R_i\Om}{1+2A_{i,i}\Om+M^2_{i,i}\Om-R_i^2\Om}, 
    \label{eq_banka1} 
\\
0 & < 1-\rho_i\Om R_i\Om,  
    \label{eq_R_i} \\
    {\mbox{where }} &~\quad
R_i\Om  =\sum_{j\neq i}M_{j,i}\Om .
\label{eq_radius_G_eig} 
    \end{align}
\end{lemma}

\vspace{-0.2in}
\begin{pf}
The conditions of this lemma are used to show that 
the eigenvalues of the contraction gain $G\Om$ have magnitude less than one, and therefore, the MIMO ILC convergence condition in Eq.~\eqref{eq_contraction_gain_condition} is met. By the Geršgorin theorem~\cite{horn2012matrix},  all the eigenvalues of the contraction gain $G\Om$ are in the union of Geršgorin discs centered at $C_i\Om = G_{i,i}\Om$ with radius $R_i\Om=\sum_{j\neq i}|G_{j,i}\Om|$. 
Then, the eigenvalues of the contraction gain $G\Om$ are less than one if all the Geršgorin discs are bounded by the unit circle centered at the origin, i.e., 
\begin{align}
|C_i\Om| + R_i\Om  < 1, 
\label{eq_gen_condition_based_on_radius}
\end{align}
which can be rewritten using Eq.~\eqref{eq_contraction_gain} 
as 

\vspace{-0.1in}
{\small{ 
\begin{align}
    |1-\rho_i\Om-\rho_i\Om\Delta_{i,i}\Om|&<1-\sum_{j\neq i}|\rho_i\Om\Delta_{j,i}\Om|.
    \label{col_wise}
\end{align}
}}

\vspace{-0.1in}
Since the left hand side (LHS) of  Eq.~\eqref{col_wise} is nonnegative, and needs to be strictly less than the right hand side (RHS), 
the RHS is required to be positive, which is satisfied due to the  condition in Eq.~\eqref{eq_R_i} and  $\rho_i\Om$ being positive from Eq.~\eqref{eq_banka1}.
Since both sides of  Eq.~\eqref{col_wise} are nonnegative, squaring them and using Eq.~\eqref{eq_unknownModelUncertainty_def} results in 
\begin{align*}
    (1-\rho_i\Om-\rho_i\Om A_{i,i}\Om)^2 &+ (\rho_i\Om B_{i,i}\Om)^2 \\
         &< (1-\rho_i\Om R_i\Om)^2. 
\end{align*}
Expanding the squares, and rearranging, yields 
\begin{align}
    \rho_i^2\Om & \left[ 
    1  +2A_{i,i}\Om   +M_{i,i}^2\Om -R_i^2\Om \right]  
    \nonumber \\ 
    & \qquad < ~ 2\rho_i\Om
    \left[  1+A_{i,i}\Om     -R_i\Om     \right]. \label{eq_temp}
\end{align}
Since the radius $R_i\Om$ is non-negative, 
squaring the condition in  Eq.~\eqref{eq_banka2} and using $A_{i,i}^2\Om  \le M_{i,i}^2\Om $,  
results in 
\begin{align}
 R_i^2\Om &  < 
    1+2A_{i,i}\Om + M_{i,i}^2\Om.
    \label{eq_temp_SD_1}
\end{align}
Since the iteration gain is positive $\rho_i\Om >0$ from LHS of Condition~\eqref{eq_banka1} and from Eq.~\eqref{eq_temp_SD_1}, $\rho_i\Om [1+2A_{i,i}\Om + M_{i,i}^2\Om -R_i^2\Om ] $ is positive and can be divided from both sides of Eq.~\eqref{eq_temp} to obtain  
\begin{align}
    \rho_i\Om  <  \frac{  2 \left[ 1+A_{i,i}\Om     -R_i\Om \right]}{ %
    1  +2A_{i,i}\Om   +M_{i,i}^2\Om -R_i^2\Om}
    = P\Om,   
     \label{eq_temp_2}
\end{align}
which is satisfied due to the condition in Eq.~\eqref{eq_banka1}. 
Thus, the conditions of the lemma ensure that the  contraction gain $G\Om$ has eigenvalues with magnitude less than one based on  Eq.~\eqref{eq_gen_condition_based_on_radius}. 
\qed \end{pf}

\vspace{-0.1in}
\begin{remark}
\label{conser1_require_review}
The conditions on the iteration gain $\rho\Om$ in Lemma~\ref{Lemma_convergence_conditions} are only sufficient and not necessary, and therefore, are conservative and may not result in the fastest possible convergence. Nevertheless, they ensure convergence to exact tracking. 
\end{remark}
\vspace{-0.05in}
\begin{lemma}[Bounded-uncertainty convergence]
\label{lem_conserative_rho}
When each component $\Delta_{j,l}\Om$ in model uncertainty  is bounded in magnitude as in Eq.~\eqref{eq_bound_uncertainty_computation}, i.e.,
\begin{equation}
    M_{j,l}\Om< \overline{\Delta}_{j,l}\Om,
    \label{eq_con_cons_lem3}
\end{equation}
the MIMO ILC convergence conditions in Lemma~\ref{Lemma_convergence_conditions} are satisfied if 
\begin{align}
    0<\rho_i\Om & < \overline{\rho}_i\Om
       \label{eq_rho_conser} \\
    \rho_i\Om \overline{\Delta}_{R,i}\Om & <
   1,
    \label{eq_rho_conser_2}  \\
\overline{\Delta}_{R,i}\Om & < ~ 1 - \overline{\Delta}_{i,i}\Om, 
\label{cons_cond_3_lem3} 
\end{align}
where
$\overline{\Delta}_{R,i}\Om =\sum_{j\neq i}\overline{\Delta}_{j,i}\Om$ and
\begin{equation}
\label{two_candi}
\overline{\rho}_i\Om ~=\min_{p=\pm 1} 
\frac{2 \left[ 1 +p\overline{\Delta}_{i,i}\Om - \overline{\Delta}_{R,i}\Om \right]}{1+2p\overline{\Delta}_{i,i}\Om + \overline{\Delta}_{i,i}^2\Om-\overline{\Delta}_{R,i}^2\Om} .
\end{equation}
\end{lemma}

\vspace{-0.1in}
\begin{pf}
The radius 
$R_i\Om < \overline{\Delta}_{R,i}\Om$ from  Eq.~\eqref{eq_con_cons_lem3}. Therefore, (i)~Eq.~\eqref{eq_rho_conser_2} implies  
$R_i\Om\rho_i\Om < 1$ and that  the  condition in Eq.~\eqref{eq_R_i} is satisfied
since $\rho_i\Om > 0$, and
(ii)~Eq.~\eqref{cons_cond_3_lem3} implies $R_i\Om < 1 - \overline{\Delta}_{i,i}\Om < 1 + A_{i,i}\Om$ and that the condition in 
Eq.~\eqref{eq_banka2} is met 
since $ \overline{\Delta}_{i,i}\Om > M_{i,i} ~ \ge |A_{i,i}\Om | $. 
Moreover, the upper bound on the iteration gain in 
Eq.~\eqref{eq_banka1}, denoted by $P\Om$ as in Eq.~\eqref{eq_temp_2}, is shown below to be a monotonic  function  of each variable $R_i\Om \in [0, \overline{\Delta}_{R,i}\Om]$, $M_{i,i}\Om\in [0,\overline{\Delta}_{i,i}\Om]$ and $A_{i,i}\Om\in [-\overline{\Delta}_{i,i}\Om,\overline{\Delta}_{i,i}\Om]$.
The upper bound $P\Om$ decreases with increasing $M_{i,i}$ and is therefore minimized at $M_{i,i}= \overline{\Delta}_{i,i}\Om$. With $X=1+A_{i,i}\Om$ and $Y=1+2A_{i,i}\Om+M_{i,i}^2$, 
\begin{equation*}
\begin{split}
\frac{\partial P}{\partial R_i\Om}
&= 2\frac{-(Y-R_i^2\Om)-(X-R_i)(-2R_i\Om)}{(Y-R_i^2\Om)^2}\\
&= 2\frac{-(R_i\Om-X)^2 -(Y-X^2)}{(Y-R_i^2\Om)^2} \le 0
\end{split}
\end{equation*}
since $Y-X^2\ge 0$ as $M_{i,i}\Om\ge A^2_{i,i}\Om$.  Therefore, for independent of $A_{i,i}\Om$ and $M_{i,i}\Om$, $P\Om$ is minimized when $R_i\Om = \overline{\Delta}_{R,i}\Om$.
Finally, with  $X'=1-\overline{\Delta}_{R,i}\Om$ and $Y'=1-\overline{\Delta}_{R,i}^2\Om+\overline{\Delta}_{i,i}^2\Om$, 
\begin{equation*}
\begin{split}
\frac{\partial P}{\partial A_{i,i}\Om}
&=2\frac{(Y'+2A_{i,i}\Om)-2(X'+A_{i,i}\Om)}{(Y'+2A_{i,i}\Om)^2} \\
&=2\frac{Y'-2X'}{(Y'+2A_{i,i}\Om)^2},
\end{split}
\end{equation*}
which does not change sign. 
Consequently, the smallest $P\Om$ occurs at either $A_{i,i}\Om=\pm \overline{\Delta}_{i,i}\Om$
and thus, 
satisfying 
Eq.~\eqref{eq_rho_conser} ensures that Eq.~\eqref{eq_banka1} is satisfied.
\qed \end{pf}

\begin{remark}
\label{rem_6_gain_selection}
The MIMO ILC converges from Lemma~\ref{lem_conserative_rho} if (i)~the  uncertainty bounds  are sufficiently small to satisfy Eq.~\eqref{cons_cond_3_lem3}  and (ii)~the nonzero iteration gain $\rho_i\Om$is chosen to be sufficiently small to satisfy Eqs.~\eqref{eq_rho_conser} and \eqref{eq_rho_conser_2} 
for all $1 \le i \le m$. 
\end{remark}

\begin{remark}
\label{connect_prob_actual}
If bounds on the modeling error are estimated from data as in Remark~\ref{rem_uncertain_coeff_bound} with some confidence level, then satisfying the conservative conditions of Lemma~\ref{lem_conserative_rho} ensures convergence to exact tracking with at least the same level of confidence. 
\end{remark}

\begin{remark}
\label{impact_noise}
Noise in measurements can limit the achievable convergence. However, the tracking error with ILC tends to be small if the noise is small~\cite{tien2005iterative}.
\end{remark}

\subsection{ILC algorithm}\label{ilc_procedure}
The ILC design and procedure are described below, and summarized in  Algorithm~\ref{algo:mimo_ilc}.

\begin{algorithm}[!ht]
\caption{MIMO ILC through Machine Learning}\label{algo:mimo_ilc}
\textbf{1. Initialization:}
Set $I_0\Om = O_d\Om,k=0$.
Select error threshold $\epsilon$ and the maximum iteration steps $k_{max}$;\\
\textbf{2. Initial input:} k = 0. \\
Apply input $I_0\Om$ to the system and measure output $O_0\Om$;\\
\textbf{3. Perturbed input:} k = 1. \\
Apply perturbed input $I_1\Om=I_0\Om+I_p\Om$ to the system and measure output  $O_1\Om$
\\ 
\textbf{4. Model estimation:} \\
Compute $O_{j,p}\Om=O_{j,1}\Om-O_{j,0}\Om$;\\ 
Use observed input $I_p\Om$ and output $O_{j,p}\Om$ to estimate the each $S^P_{j,l}$ and model $\Se_{j,l}$ from Eq.~\eqref{E_mimo_cgpr} and \eqref{V_mimo_cgpr};
\\
\textbf{5. Iteration gain selection:}\\
Estimate  bounds $\overline{\Delta}_{j,l}$ on uncertainty $\Delta_{j,l}$ as in Eq.~\eqref{eq_bound_uncertainty_computation}; \\
Select iteraiton gain $\rho\Om$ to meet the upper bound $\overline{\rho}\Om$ on the iteration gain from Eq.~\eqref{two_candi}.
\\
\textbf{6. Iterative input correction:} $2\le k\le k_{max}$.\\
Obtain $I_2\Om=I_0\Om+\Se^{\dagger}\Om\rho\Om (O_d\Om-O_0\Om)$;\\ 
Apply  $I_2\Om$ to the system and measure  $O_2\Om$;\\
Compute tracking error $E_2\Om=O_d\Om-O_2\Om$ and its time domain representation $E_2(t)$;\\
Compute maximum tracking error  $\overline{E}_{j,2}=\max_t |E_{j,2}(t) |$; 
{\bf{while}} There exists $ j \in[1,m] $ such that $ \overline{E}_{j,k} \geq \epsilon$ 
\textbf{and} $k\leq k_{max}$
{\bf{do}}\\ $k = k + 1$; \\
Compute $I_k$ from Eq.~\eqref{eq_update_IIC} using $I_{k-1}$ and $O_{k-1}$;\\
Apply $I_k\Om$ to the system and measure $O_k\Om$;\\
Compute $E_{j,k}(t)=O_{j,d}(t)-O_{j,k}(t)$ and $\overline{E}_{j,k}$; \\
{\bf{end while}} 
\end{algorithm}

\begin{enumerate}
\item 
{\bf{Initial input:~}} 
At the initial step $k=0$, the desired output $O_d$ is applied as the reference trajectory to be tracked by the system $S$, i.e., the input $I_0$ applied to the system is selected as  
$
I_0\Om = O_d\Om,
$
with the resulting output $O_0$. The error between this initial output $O_0$ and the desired output $O_d$ is corrected through the MIMO ILC. 
\item 
{\bf{Perturbed input:~}} 
To estimate the local model at the operating point, the next input $I_1$ applied to the system at step $k=1$ is selected as the summation of $I_0$ and a small-amplitude perturbation $I_p$, i.e., 
\begin{align} 
I_1\Om= I_0\Om +{I_p}\Om
\label{eq_perturb_I1}
\end{align}
and the resulting output is 
$O_1$. 
The use of the perturbation input around the initial input $I_0$ helps to generate a linear localized model around the operating point.
Moreover, the desired output $O_d$ (and therefore the initial input $I_0=O_d$) might have low frequency content. In contrast, the perturbation input $I_p$ can be selected to be frequency rich and provide the persistence of excitation needed for model acquisition~\cite{devasia2017iterative}. 
\item
{\bf{Model estimation:~}} 
The perturbation in the output $O_p$ 
caused by the perturbation $I_p$ in the  input $I_1$, was found as 
the difference in the output in the first two ILC steps  \begin{equation}
O_p\Om = O_{1}\Om-O_{0}\Om.
\label{eq_perturb_output_def} 
\end{equation}
The  $j^{th}$ 
component of the output perturbation 
i.e.,    
\begin{equation}
O_{j,O}\Om = O_{j,p}\Om, 
\label{eq_perturb_output} 
\end{equation}
and the perturbation $I_O\Om=I_p\Om$ are used to estimate the 
each term $S^P_{j,l}$ (with $1 \le l \le n$) and the associated variance,  from Eq.~\eqref{E_mimo_cgpr} and Eq.~\eqref{V_mimo_cgpr} through multi-input system identificaiton as in Lemma~\ref{MISO_sys_id_input_weighted_kernel}
for $j^{th}$ row of $S$.  
\item
{\bf{Iteration gain selection:~}} 
The iteration gain $\rho\Om$ is selected by  estimating bounds $\overline{\Delta}_{j,l}\Om$ on the magnitude of the uncertainty $\Delta_{j,l}\Om$ using Eq.~\eqref{eq_bound_uncertainty_computation} of Lemma~\ref{Lemma_bounds_uncertainty} and then using them to select the each diagonal term $\rho_i\Om$ of the
iteration gain $\rho\Om$ to satisfy conditions in Lemma~\ref{lem_conserative_rho}.
$\rho\Om$ is set to zero if conditions of Lemma~\ref{lem_conserative_rho} can not be met or
frequency $\omega$ is beyond the desired tracking bandwidth
%
\item
{\bf{Iterative input correction:~}} 
The iterations are repeated  for $2 <  k \le k_{max}$, or till the
the maximum tracking error of any output at step $k-1$ is greater than the given error threshold $\epsilon$, i.e.,
 \begin{equation}
    \max_t|O_{j,d}(t)-O_{j,k-1}(t)| = \overline{E}_{j,k-1} \ge \epsilon,
 \end{equation}
 where the input  $I_k$ is updated based on the input $I_{k-1}$ and $O_{k-1}$ in step $k-1$ using Eq.~\eqref{eq_update_IIC}, while the input $I_2$ in step $k=2$ is updated based on $I_0$ and $O_0$ to not use the perturbed output in step $k=1$. 
\end{enumerate}

\section{Experiments}
The performance of an SEA robot were comparatively evaluated, with and without ILC.

\subsection{ILC for hole cleaning}
\subsubsection{Experimental system}
A low-profile 3-DOF robotic arm was used in the experiment to mimic pilot hole cleaning in confined spaces, as illustrated in \figurename \ref{fig_ConfigPic}. 
The joint  actuators were HEBI X5-4 series elastic actuators, and the links between the joints were PVC black pipes with diameter 
$\diameter =1.25$ inch and lengths $l_1 = 16.90$ cm and $l_2 = 17.97$ cm.  The brush (Forney 70485 Tube Brush) had a  bristle diameter $\diameter =12$ mm, and 
the  effective length to the tip of the  end-effector brush
from the center of the joint $\theta_3$  actuator was length $l_3 = 15.86$ cm as shown in Fig.~\ref{fig_ConfigPic}.
The plate in the front of the robot was drilled with evenly-spaced holes of diameter $\diameter =  6.2 $ mm to represent a part to be cleaned with the robot. A MATLAB interface with relevant HEBI libraries were used to send commands to and receive data from the robotic arm, and data processing was done with MATLAB. 
The sampling rate for the input and output were $100$ Hz. The internal feedback frequency of the SEA robot was also set as $100$ Hz.

\subsubsection{Task description} \label{exp_task_descrip}
\begin{figure}
    \centering
    \includegraphics[width=0.9\columnwidth]{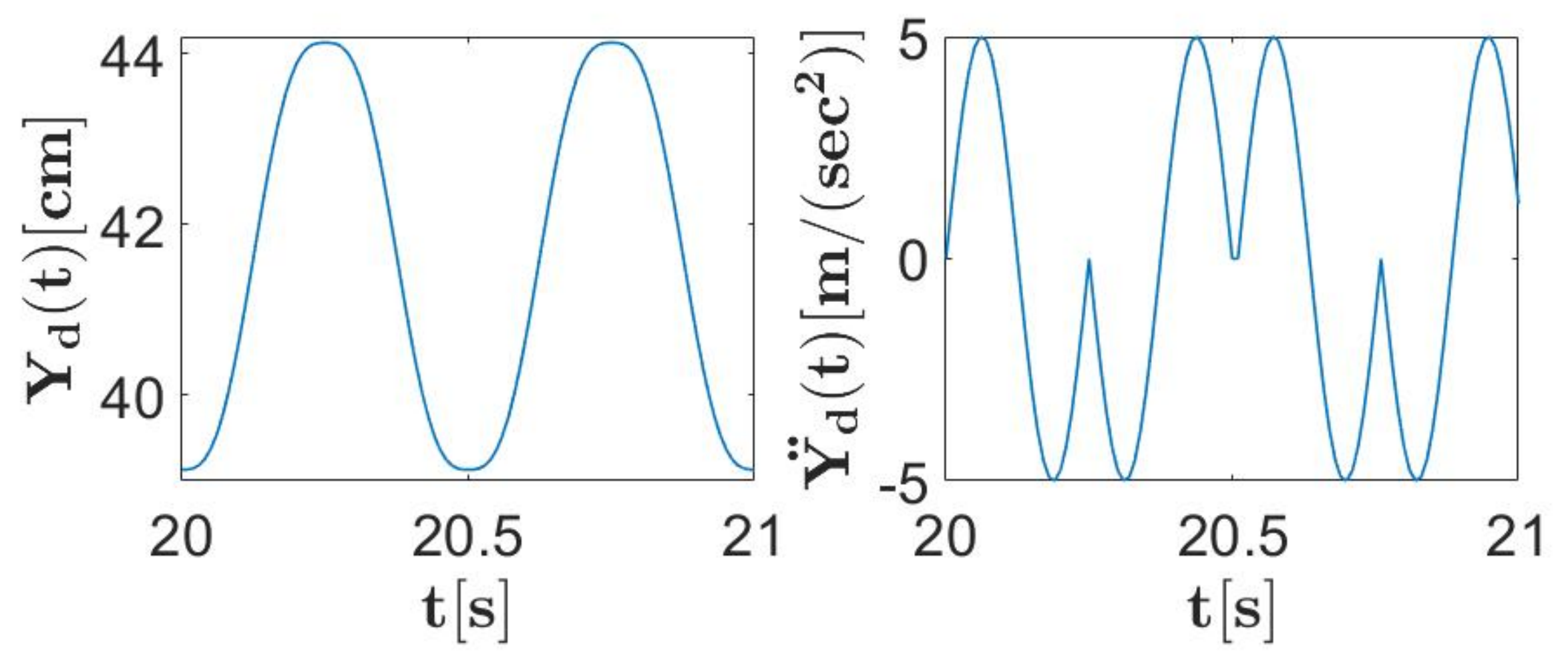}
    \caption{Desired motion $Y_d$ (left) and acceleration $\ddot{Y}_d$ (right) of the brush tip in the Y direction during $t \in [20,21]$ s.
    }
    \label{fig:Y_ProfilePic}
\end{figure}

The operation studied here is the cleaning of a single hole, which requires the robot to execute a periodic forward-and-backward  movement of the brush tip in the $Y$ direction in Fig.~\ref{fig_ConfigPic}. The desired position $Y=Y_d$ is described by its acceleration $\ddot{Y}_d$, for time $t \in [0,~ t_f]$, as 
\begin{equation}
\begin{array}{rcll} 
\ddot{Y}_d(t) 
& = & A \sin(\omega_T (t-\underline{t}_{k_c})) & 
\underline{t}_{k_c}  \le t <   {t}_{k_c} \\
& = &  \ddot{Y}_d({t}_{k_c} - (t-{t}_{k_c}) )  & 
{t}_{k_c} \le t <   \overline{t}_{k_c} \\
& = & 0  & {\mbox{otherwise}}   \end{array}
\label{eq_ydd_temp}
\end{equation}
with initial conditions $ \dot{Y}_d(0)=0, Y_d(0)=\underline{Y}$, 
where the amplitude of the acceleration is $A = \frac{8\pi d}{T^2}$ and frequency $\omega_T = \frac{4\pi}{T}$ with $T$ as the time period for each forward-and-backward motion, $d$ as the stroke length, which is kept fixed at $5$~cm in the following, and 
$\underline{Y}= l_b+l_3=39.13$ cm represents the situation when the tip of the brush is just touching the plane of the plate with holes as in Fig.~\ref{fig_ConfigPic}.
Moreover, $\underline{t}_{k_c} = t_1 + k_c T $, ${t}_{k_c} = t_1 + (k_c+0.5)T  $, and $\overline{t}_{k_c} = t_1 + (k_c+1) T  $,
with integer  $0 \le k_c < (k_N-1)$, where $k_N=20/T$ is the number of cleaning cycles,  $t_1=20$~s is the amount of initial and final period without motion before and after the cleaning cycles, and the final time is $t_f = 2*t_1 +k_N T$.
An example trajectory $Y_d$ with time period $T=0.5$ s is shown in Fig.~\ref{fig:Y_ProfilePic}.
The desired  position $X = X_d$ of the brush tip is at the center of the hole to be cleaned, and the brush is to be held perpendicular to the plate with the holes, i.e., the angle $\Theta$ in Fig.~\ref{fig_ConfigPic} is to be kept constant at the desired value $\Theta_d = \pi/2$~rad. 

\subsubsection{System input and output}
\begin{figure}
\begin{center}
\includegraphics[width=\columnwidth]{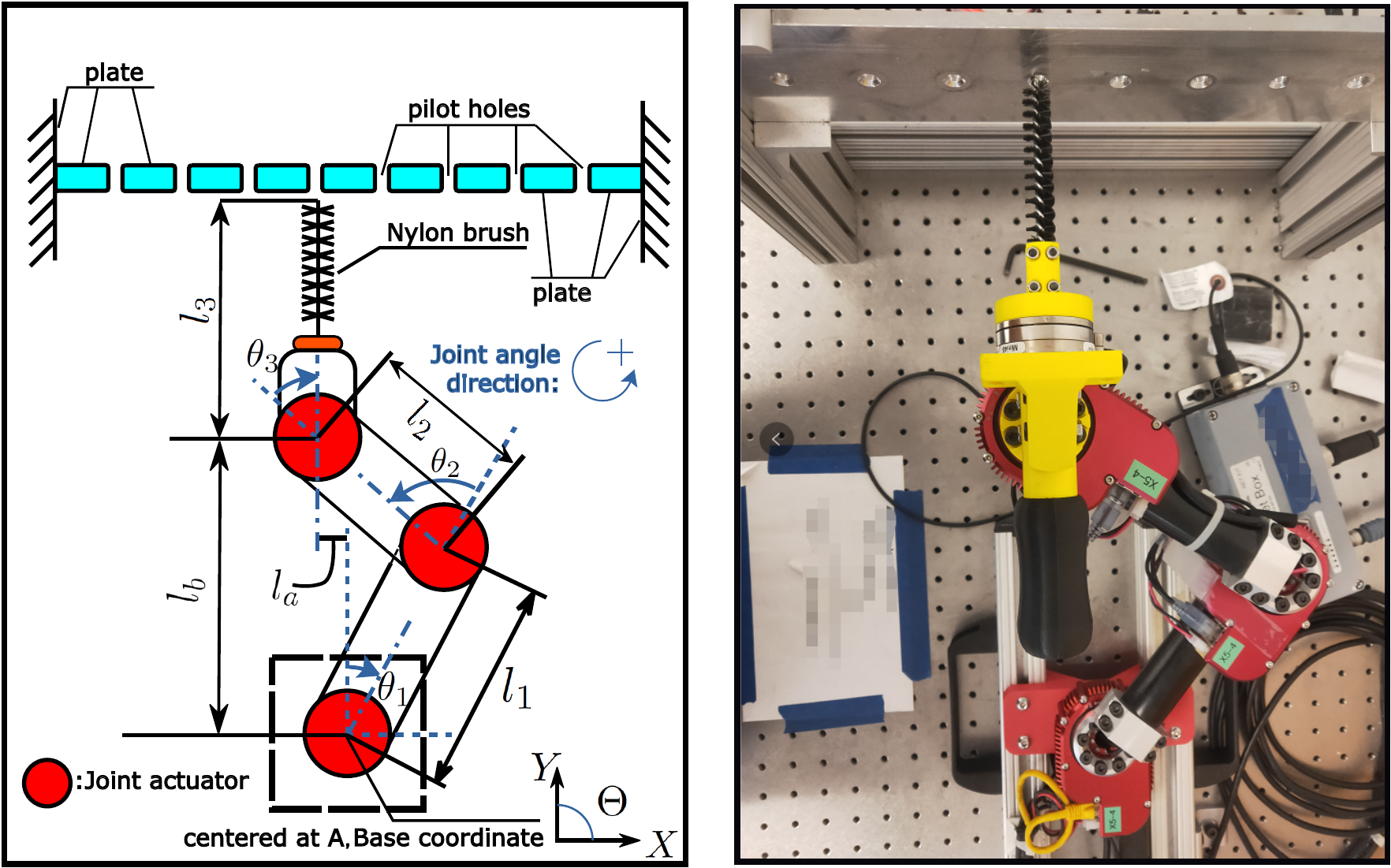}
\caption{Schematic drawing (left) and top view (right) of the experimental SEA robot. The cleaning task for a specific pilot hole consists of letting the brush achieve a periodic forward-backward motion with stroke length $d$, which should be perpendicular to the plate, i.e, end-effector orientation $\Theta = \pi/2$ rad. The controlled output are the local joint angles $\theta_1, \theta_2, \theta_3$. The pose shown in figure depicts the initial pose at the start of  the hole-cleaning task.}
\label{fig_ConfigPic}
\end{center}
\end{figure}

The controlled output in the experimental system were the local joint angles 
$O=[\theta_1,\theta_2,\theta_3]^T$ as in Fig.~\ref{fig_ConfigPic}, and the control input $I$ were the reference joint angles $I=[\theta_{1,r},\theta_{2,r},\theta_{3,r}]^T$ applied to the  feedback-based  controllers at each joint. The brush tip trajectory $X, Y, \Theta$ are related to the output $O$ as 
\begin{align}
Y(t)&=l_1\cos (\Phi_{1}(t))+l_2\cos(\Phi_{2}(t))+l_3\cos(\Phi_3(t)),\label{eq_tran_1}\\
X(t)&=-l_1\sin (\Phi_1(t))-l_2\sin(\Phi_2(t))-l_3\sin(\Phi_3(t)),\label{eq_tran_2}\\
\Theta(t)&=\Phi_3(t)+\pi/2, 
\label{eq_tran_3}
\end{align}
where $\Phi_k(t)=\sum_{i=1}^k\theta_i(t)$.
Consequently, the desired output $O_d$ (i.e., the desired joint angles $\theta_{j,d}$, $1\le j\le 3$) can be obtained from the known desired tip position $X_d, Y_d$ and orientation $\Theta_d$, as 
\begin{align}
\theta_1(t)&=-\arctan{\left(\frac{-l_a}{l_b+l_s(t)}\right)}\nonumber\\
&\qquad\qquad-\arccos{\left(\frac{(l_b+l_s(t))^2+l_a^2+l_1^2-l_2^2}{2l_1\sqrt{l_a^2+(l_b+l_s(t))^2}}\right)}, \label{direct_comp_the_1}\\
\theta_2(t)&=\pi-\arccos{\left(\frac{l_1^2+l_2^2-(l_a^2+(l_b+l_s(t))^2)}{2l_1l_2}\right)}, 
\label{direct_comp_the_2}\\
\theta_3(t)&=-(\theta_1(t)+\theta_2(t)), 
\label{direct_comp_the_3}
\end{align}
where $l_s(t)=Y(t)-\underline{Y}$, $l_b=l_1\cos{(\Phi_1(0))}+l_2\cos{(\Phi_2(0))}$ and $l_a=|-l_1\sin{(\Phi_1(0))}-l_2\sin{(\Phi_2(0))}|$, 
and the initial pose of the robot
in Fig.~\ref{fig_ConfigPic} yields 
$\Phi_1(0)=-0.6756$~rad, 
$\Phi_2(0)=1.0007$ rad, 
and
$\Phi_3(0)=0$ rad. 
Finally, given the initial pose of the robot, $\theta_1$ is always negative for the specific hole to be cleaned.

\subsubsection{Need for ILC}
\begin{figure}[!t]
    \centering
    \includegraphics[width=\columnwidth]
    {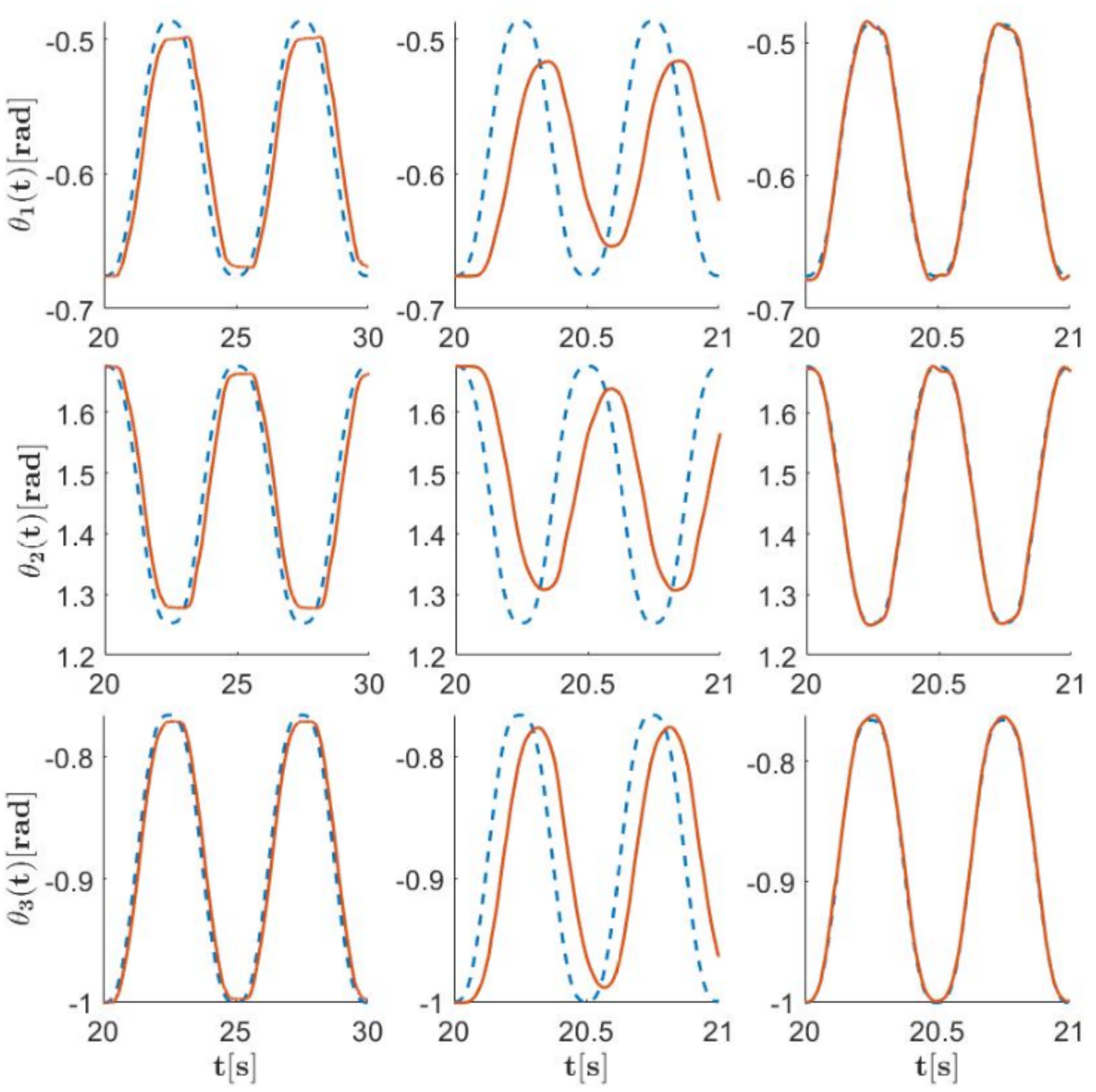}
    \caption{Comparison of desired output $O_d$ (dashed line) and achieved output $O$ (solid line) with and without ILC for three cases: (left) slower trajectories with time period $T=5$~s without ILC; (middle) faster trajectories with time period $T=0.5$ s without ILC; and (right) faster trajectories with time period $T=0.5$ s with ILC.}
    \label{fig:slow_fast_ttt_compare}
\end{figure}

\begin{figure}[!t]
    \centering
    \includegraphics[width=0.8\columnwidth]{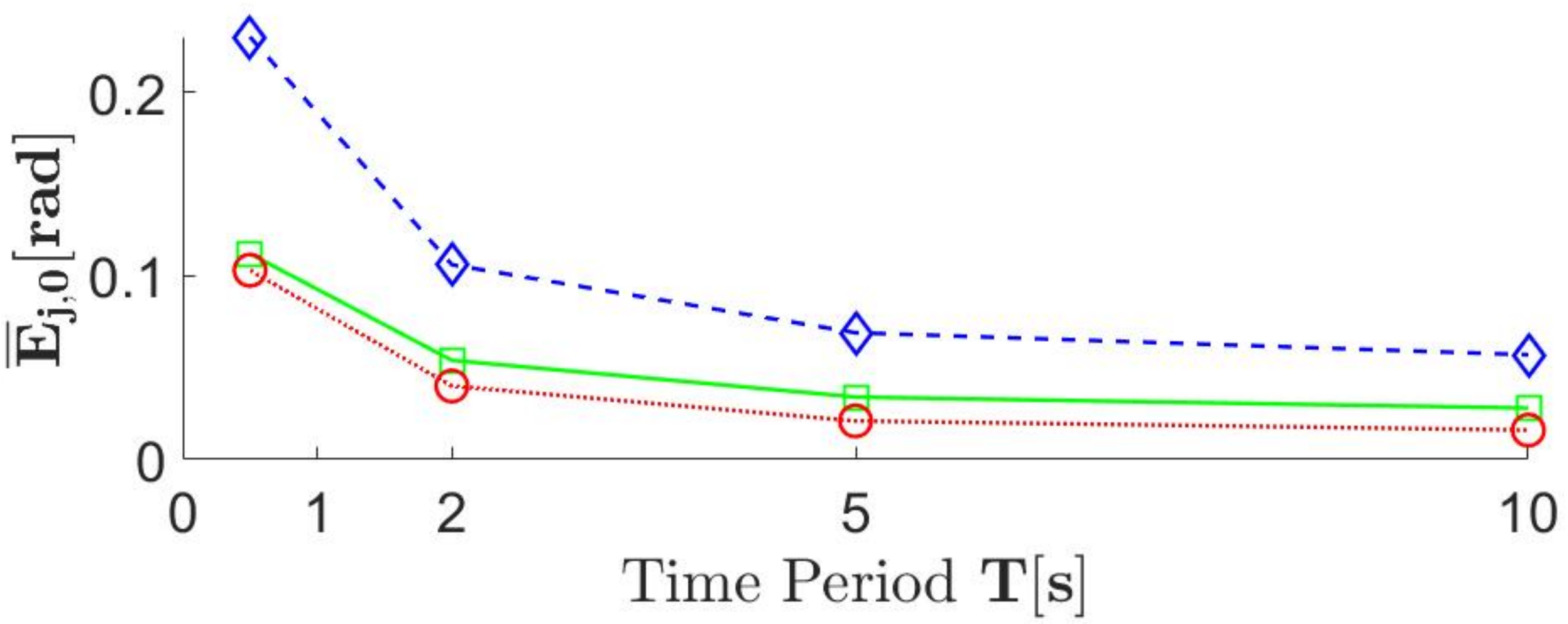}
     \caption{Joint-tracking error $\overline{E}_{j,0}$ defined in Algorithm~\ref{algo:mimo_ilc} increases as time period $T$ decreases, i.e., for faster cleaning motion: 
      $\overline{E}_{1,0}$ (square), $\overline{E}_{2,0}$ (diamond) and $\overline{E}_{3,0}$ (circle).}
    \label{fig:err_diffExeTime_plot}
\end{figure}

If the brush is to be moved slowly (with large time period $T$), then the robot's joint controllers can successfully track the reference joint angles  with sufficient precision, i.e.,
with the reference input $I = O_d$ at iteration step $k=0$, the achieved output $O$ is close to the  desired output trajectory $O \approx O_d$. However, if the brush is moved vigorously (with small time period $T$), then the tracking is not as good, as seen in Fig.~\ref{fig:slow_fast_ttt_compare}. Note that as the operation speed increases,
the output $O$, i.e., joint angles $\theta_{j,0}$ (at initial iteration step)
do not follow the desired joint angles $\theta_{j,d}$. In particular, as time period $T$  decreases, the maximum value of the joint-tracking error $\overline{E}_{j,0}$
increases by $ 4 $ times from $ \overline{E}_{1,0}= 0.028$ rad, $ \overline{E}_{2,0}=0.057$ rad,
$ \overline{E}_{3,0}=0.016$ rad at time period $T=10$ s to $ \overline{E}_{1,0}= 0.112$ rad, $ \overline{E}_{2,0}=0.230$ rad,
$ \overline{E}_{3,0}=0.103$ rad
at time period $T=0.5$ s as seen in Fig.~\ref{fig:err_diffExeTime_plot} and quantified in  Table~\ref{tab:e_bar_jionts}.

While the brush is flexible enough to handle some distortion, repeated 
large errors in the positioning in the $X$ direction and  in the orientation angle $\Theta$ can damage the brush.
Therefore, ILC in Eq.~\eqref{eq_update_IIC} is used to improve the positioning precision by correcting for motion-induced errors
in the desired output $O_d$ with time period $T=0.5$~s as in Fig.~\ref{fig:slow_fast_ttt_compare}.  Note that the ILC procedure includes steps to find local models (that includes contact effects) for each hole cleaning task. Both the modeling, and the iterative corrections have to  be repeated for holes that are far from each other if there is substantial change in robot pose. Nevertheless, an advantage in the proposed application is that the ILC  can be carried out ahead of time, outside of the confined space, provided the pose and support of the robot is similar when placed in the confined space.

\suppressfloats

\suppressfloats
 
 \begin{table}[] \centering \caption{Impact of faster cleaning motion (smaller time period $T$)  on joint-tracking error $\overline{E}_{j,0}$ defined in Algorithm~\ref{algo:mimo_ilc}.} \begin{tabular}{|c|c|c|c|}        \hline Time Period, $T$[s] & $\mathbf{\overline{E}_{1,0}}$ [rad] & $\mathbf{\overline{E}_{2,0}}$ [rad] & $\mathbf{\overline{E}_{3,0}}$ [rad] \\ \hline 0.5 &  0.112 & 0.230 & 0.103   \\[-0.05in] 2   & 0.054 & 0.106 & 0.040   \\[-0.05in] 5  & 0.034 & 0.069 &  0.021  \\[-0.05in] 10  & 0.028 & 0.057 & 0.016 \\ \hline \end{tabular} \label{tab:e_bar_jionts} \end{table}   

\subsection{ILC methods}
The MIMO ILC experiments
followed Algorithm~\ref{algo:mimo_ilc} 
to correct positioning errors during fast cleaning, with time period $T=0.5$ s.

\subsubsection{Initial input}
\begin{figure}[!t]     \centering     \includegraphics[width=0.8\columnwidth]{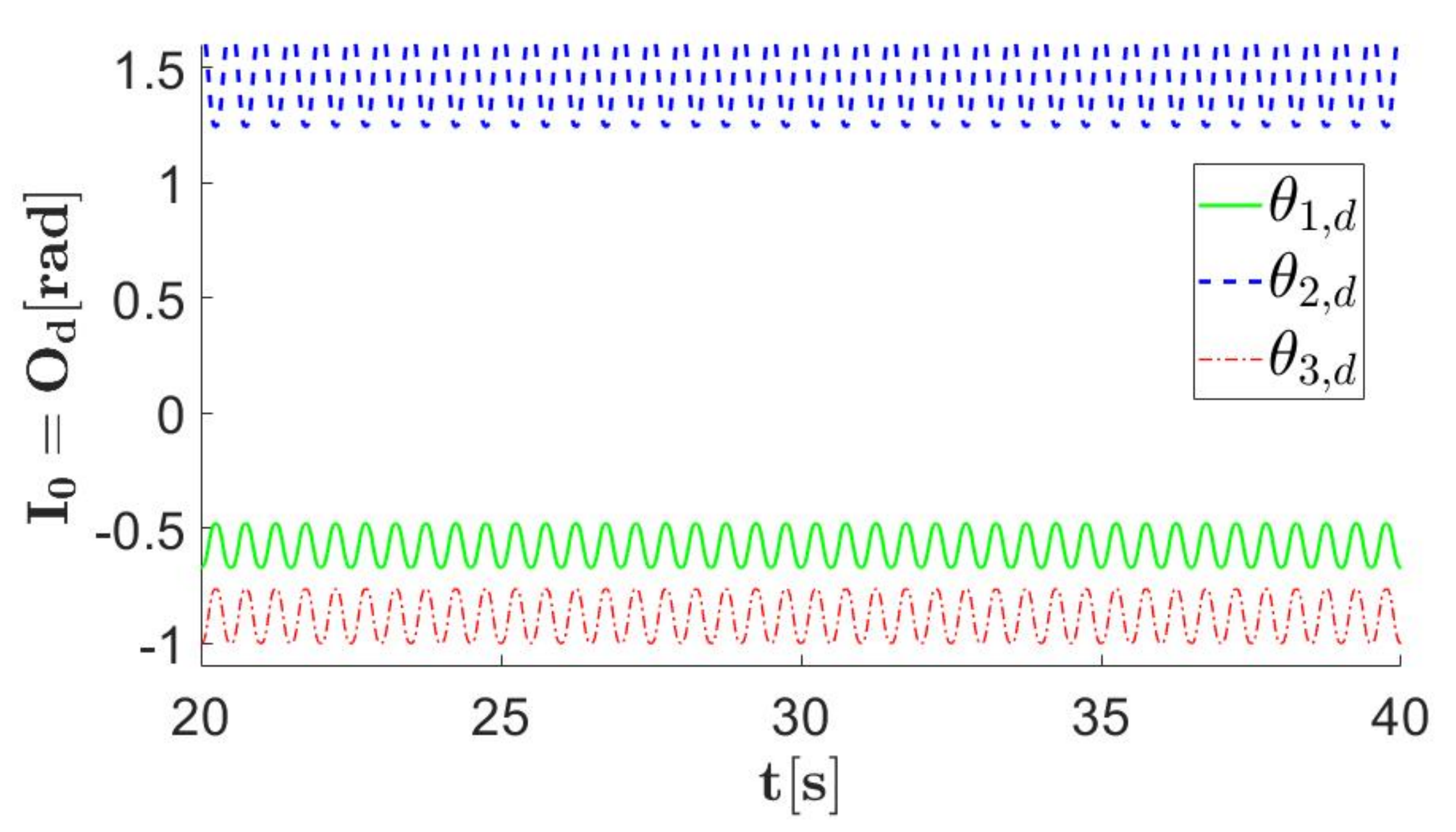}     \caption{The desired output $O_d$, i.e., desired joint angles      $[\theta_{1,d},\theta_{2,d},\theta_{3,d}]^T$, which are held constant outside the shown time interval at
 $[\theta_{1,d}(0),\theta_{2,d}(0),\theta_{3,d}(0)]^T=[-0.6756,1.6763,-1.0007]^T$rad.    }    \label{fig:Od_I0_timePlot} \end{figure}

In the initial ILC step $k=0$, the desired output $O_d$ was selected as the desired joint angles $\{\theta_{j,d}\}_{j=1}^3$ computed from the known desired brush-tip trajectory $\{X_d,Y_d,\Theta_d\}$ using Eqs.~\eqref{direct_comp_the_1} to \eqref{direct_comp_the_3}, as shown in 
Fig.~\ref{fig:Od_I0_timePlot}. The initial input $I_0=O_d$ was applied to the SEA robot and the output $O_0$ was measured. 

 \suppressfloats 

\subsubsection{Perturbed input and model estimation}
\label{subsub:systemEstimate}
\begin{figure}[!t]
    \centering
    \includegraphics[width=\columnwidth]{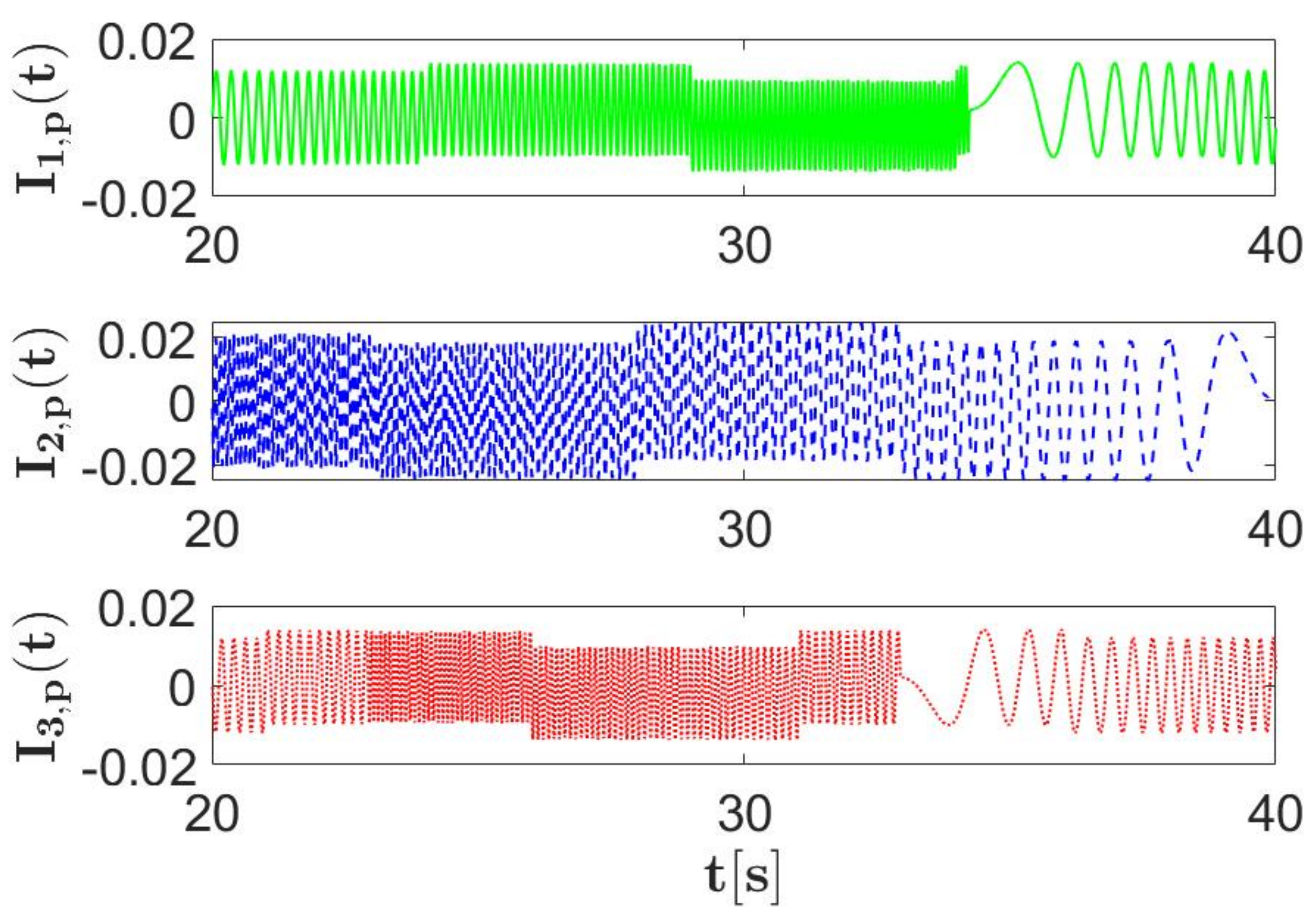}
    \caption{
    Input perturbation $\{I_{l,p}\}_{l=1}^3$  at ILC step $k=1$ with a mixture of chirp and staircase signals  were added at ILC step $k=1$. 
    The input perturbation $I_p$ was zero outside the shown time interval.
    }
    \label{fig:excitationChirp}
\end{figure}

The   input perturbation $I_p$ in ILC step $k=1$ can be selected to be frequency rich and provide the persistence of excitation needed for model acquisition~\cite{devasia2017iterative}.
$O_{j,p}$ represents the output caused by the input perturbation $I_p$. For the experiments, the input  perturbation$I_p$
was chosen to be the sum of  chirp signals ($C_p$) and staircase functions ($H_p$),  with different patterns 
$ I_{j,p}=C_{j,p}+H_{j,p}$ for each joint $j$, as illustrated in \figurename~\ref{fig:excitationChirp}. 
The chirp functions were, for time $ t\in [0, 60]~s$ and frequency $\omega_c =0.3$~hz,
\begin{equation*}
    C_{1,p}(t)=
    0.012 \sin (2\pi \omega_c t_{1,c}^2),  \quad \forall ~20\le t\le 40
\end{equation*}
and zero otherwise, where $t_{1,c}= \mod (t-20+\sqrt{100/3}, 20)$,
with $t \in [0,60]$~s  for joint $1$. For joint 2, 
\begin{equation*}
    C_{2,p}(t)=
    0.022\sin (2\pi \omega_c (40-t)^2),  \quad \forall 20\le t\le 40
\end{equation*}
and zero otherwise, and for joint 3, 
\begin{equation*}
C_{3,p}(t)=
\begin{cases}
-0.012\sin (2\pi \omega_c (t_{3,c}-10)^2)
& 10\le t_{3,c} \\
0.012\sin (2\pi \omega_c (20-t_{3,c})^2)
& t_{3,c}< 10\\
0 & \text{otherwise},
\end{cases}
\end{equation*}
where $t_{3,c}=\mod (t-30+\sqrt{50}, 20)$,
and the staircase functions consist of three consecutive 5-second steps starting from $t=t_H$ with magnitude equaling to $h_a, h_b$ and $h_c$, and are zero otherwise. The  parameters for each staircase function $H_{j,p}$ are tabulated in Table~\ref{tab:H_param}.

\begin{table}[]
\centering
\caption{Parameters for staircase functions $H_p$.}
\begin{tabular}{|c|c|c|c|c|}
\hline
staircase index & 
$t_H$ [s] & $h_a$ [rad] & $h_b$ [rad] & $h_c$ [rad] \\
\hline
$H_{1,p}$ & 24 & +0.002 & -0.002 & +0.002 \\[-0.05in]
$H_{2,p}$ & 23 & -0.003 & +0.003 & -0.003 \\[-0.05in]
$H_{3,p}$ & 21 & +0.002 & -0.002 & +0.002 \\[-0.02in]
\hline
\end{tabular}
\label{tab:H_param}
\end{table}   
From linearity, the perturbation input-output relation was, from Eq.~\eqref{eq_model_freq_response_abstract}, $O_{p}\Om= S\Om I_p\Om$. 

\begin{figure*}[!t]
  \includegraphics[width=\textwidth]{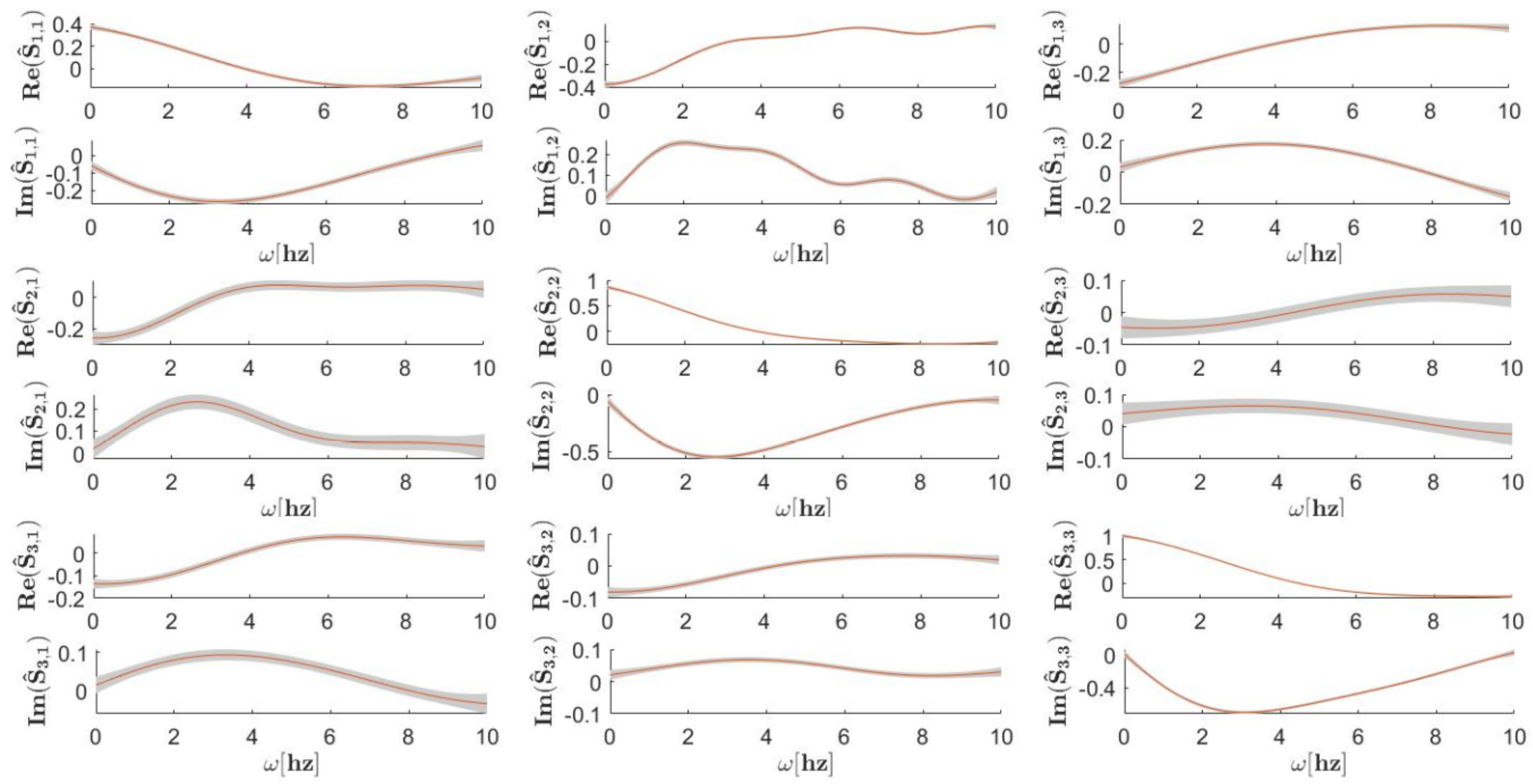}
  \caption{Bode frequency-response plots.
  Estimated model $\hat{S}$ of the system $S$  defined in Eq.~\eqref{eq_model_freq_response_abstract}. The red lines are the expected values from Eq.~\eqref{E_mimo_cgpr} and deviation of  $\pm \mathbb{V}_{j,l}\Om$ shown in gray, with the variance $\mathbb{V}_{j,l}\Om$ defined in Assumption~\ref{GPR_assumptions}.}
  \label{fig:G_overall}
\end{figure*}

With the system $S$ replaced by the Gaussian process $S^P$ as in Eq.~\eqref{eq_output_MISI_process},
the observed perturbation  $O_O$
for each joint angle $1 \le j \le 3$, i.e., 
$O_{j,O} = O_{j,p} $ 
along with the input $I_p$ were 
used to estimate the model
subsystems $\Se_{j,l}$ (with $1 \le l \le 3$)
and the associated variance $\mathbb{V}_{j,l}$,  from Eq.~\eqref{E_mimo_cgpr} and Eq.~\eqref{V_mimo_cgpr}, through the input-weighted complex kernel as in Lemma~\ref{MISO_sys_id_input_weighted_kernel}. The necessary Fourier transforms and inverse Fourier transforms were computed in MATLAB. 
The SISO kernel $\hat{k}_{j,l}$
was selected as
$     \hat{k}_{j,l}(\omega_1,\omega_2)=\sigma_{f,j,l}^2
    \exp{(-\frac{1}{2}(\omega_1-\omega_2)^*l_{j,l}^{-2}(\omega_1-\omega_2))},
    $
where $\sigma_{f,j,l}$ and $l_{j,l}$ denote the output variance and length scale, respectively. 
Then, the estimated subsystems $\hat{S}_{j,l}$ and their variance $\mathbb{V}_{j,l}$ are shown in \figurename \ref{fig:G_overall}.

\suppressfloats

\subsubsection{Iteration gain selection}\label{subsub:iterationGain_chose}
\begin{figure}[!t]
    \centering
    \includegraphics[width=\columnwidth]{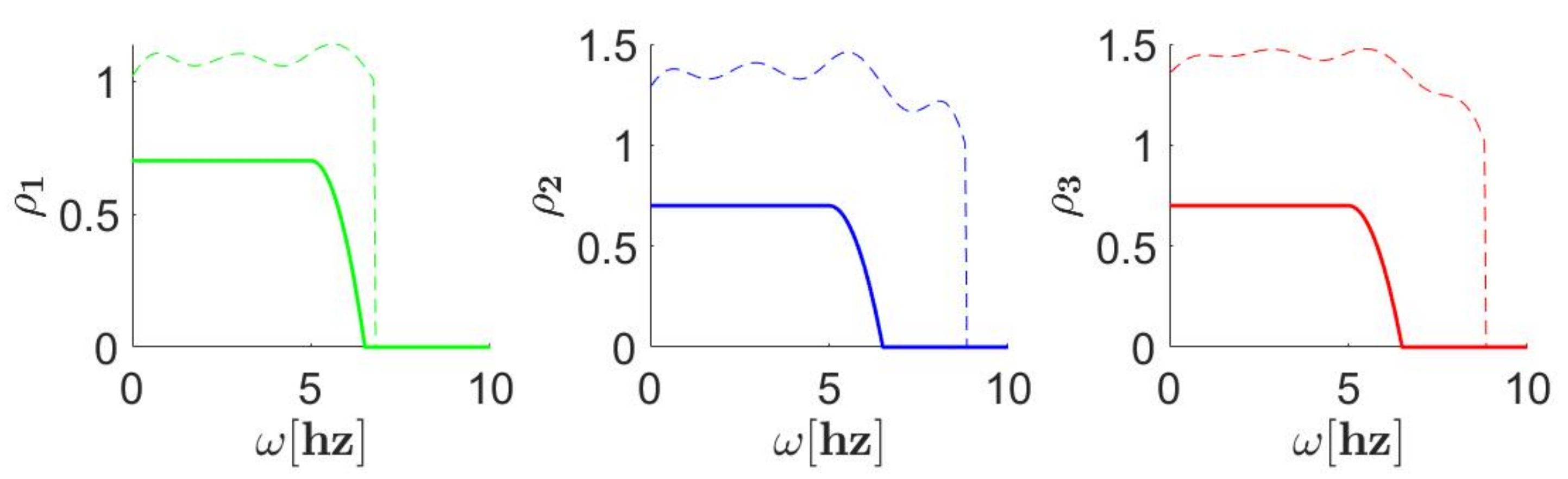}
     \caption{Selected iteration gain  $\{\rho_i\Om\}_{i=1}^3$ (solid line)  and upper bound $\{\overline{\rho}_i\Om\}_{i=1}^3$(dashed line) from Eq.~\eqref{two_candi}. 
     }
    \label{fig:iteration_gain_plot}
\end{figure}

The iteration gain $\rho\Om$ was selected to ensure ILC convergence based on the estimated model and uncertainty. 
Bounds $\overline{\Delta}_{j,l}\Om$ on the model uncertainty $\Delta_{j,l}\Om$ were obtained 
from Eq.~\eqref{eq_bound_uncertainty_computation} of Lemma~\ref{Lemma_bounds_uncertainty}, 
with 
$\gamma_{\delta}=3$ in Eq.~\eqref{ineq_bound_mag} of  Remark.~\ref{rem_uncertain_coeff_bound}. The iteration gains $\rho_i\Om$ ($i=1,2,3$) were chosen to be 
$0.7$ for $0 \le \omega \le 5$. 
Moreover, since the the desired output $O_d$ did not have significant frequency content beyond $6$~Hz, the iteration gains were reduced to zero after $6.5$~Hz, as $\rho_i\Om = \rho_i(5)(1-\frac{\omega-5}{1.5})^2$ for $5 < \omega \le 6.5$.
The upper bound $\overline{\rho}_i\Om$ and the selected iteration gain $\rho_i\Om$ are shown in  Fig.~\ref{fig:iteration_gain_plot}.

\subsubsection{Iterative input update}
At each iteration step $k \ge  3$, the error $E_{k-1} = O_d(t)-O(t)$ 
during the active cleaning period ($t \in [20, 40] s$) was computed using Fourier transform in MATLAB, and used to update the input $I_{k-1}$ to find the new input $I_k$.
For iteration step $k=2$, the new input $I_2$ was updated based on input $I_0$ and error $E_0$. 
Prior to the Fourier transform, the initial and final settling of the closed-loop controllers beyond the cleaning cycle were removed in all iterations by padding the error signal in time $E_{k-1}(t)$ with zeros before and after the end of the cleaning cycles for $5$ s and thereby, the input $I_{k}$ was updated over the time interval $t \in [15, 45] s$.

\subsection{Results \& Discussion}
\begin{figure}[!t]
    \centering
    \includegraphics[width=\columnwidth]{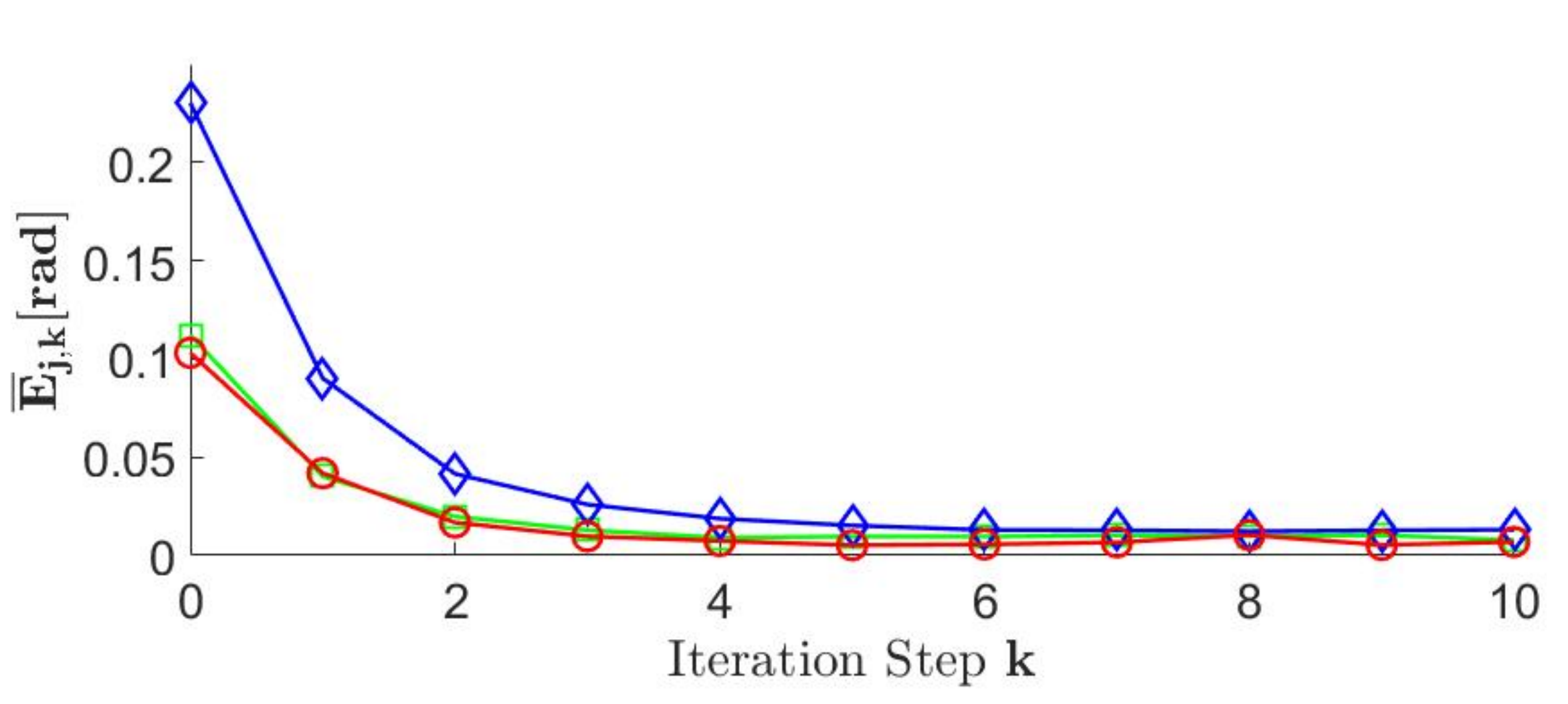}
     \caption{Reduction of joint error $\overline{E}_{j,k}$  
     with
     iteration step $k$:  $\overline{E}_{1,k}$ (square), $\overline{E}_{2,k}$ (diamond) and $\overline{E}_{3,k}$ (circle). 
     }
    \label{fig:iteration_err_plot}
\end{figure}

The ILC led to improvement in the positioning precision of the brush with the SEA robot, even in the presence of significant contact effects. The reduction of the
joint tracking error $\overline{E}_{j,k}$, 
with iteration step $k$ is shown in \figurename~\ref{fig:iteration_err_plot} and the tracking results are shown in
\figurename~\ref{fig:slow_fast_ttt_compare}.

\suppressfloats

\suppressfloats

The joint tracking error decreased from initial values of $\overline{E}_{1,0} = 0.112$~rad, $\overline{E}_{2,0} =0.230$~rad, $\overline{E}_{3,0} = 0.103$~rad at iteration step $k=0$ to final values of $\overline{E}_{1,10} = 0.008$~rad, $\overline{E}_{2,10} = 0.013$~rad, $\overline{E}_{3,10} = 0.007$~rad at iteration step $10$. The final tracking errors were close to the repeatability of the system - the non-repeatable errors in the joint positioning of the robot were experimentally estimated to be $0.004$ rad at joints 1 and 3, and $0.007$ rad at joint 2. Thus, the ILC approach led to substantial reduction of 
92\% in joint $\theta_1$, 94\% in joint $\theta_2$ and 93\% in joint $\theta_3$ in the tracking error.

An alternate approach to reduce the tracking error, without ILC, is to slow down the cleaning motion. In particular, with a time period $T=5$ s, the tracking error without ILC was $\overline{E}_{1,0} = 0.034$~rad, $\overline{E}_{2,0} = 0.069$~rad, $\overline{E}_{3,0} = 0.021$~rad. This is still larger than the 
final tracking error with ILC with a  time period $T=0.5$ s, as seen by 
comparing  the desired and actual output joint angles for the time period $T=5$ s without ILC in Fig.~\ref{fig:slow_fast_ttt_compare}.
Thus, the ILC enables at least 10-times increase in the operating speed for similar positioning precision with the SEA robot.

\section{Conclusion}
This work shows that the proposed complex-kernel Gaussian process regression  with a 
proposed  input-weighted kernel can sufficiently capture the model of a robot with series elastic actuators for precision operations even in the presence of contact effects, which in general are  challenging to model a priori. Experimental results showed more than an order increase in operating speed and around 90\% improvement in the positioning precision. Additionally, the work developed theoretical conditions to ensure convergence of an iterative learning controller for multi-input multi-output systems. 
However, the proposed  approach is only valid locally around an operating point where the error caused by nonlinearity is sufficiently small (e.g., for local cleaning operations as demonstrated in the paper), and is not suitable for large-range motions with substantial nonlinearity.  Our ongoing efforts are aimed at developing data-enabled methods to model and correct for such robot-pose-dependent nonlinearities.

\begin{ack}
This work was supported by NSF Grant CMMI 1824660. 
\end{ack}


\end{document}